\documentclass[11pt]{article}

\usepackage{amsmath}
\usepackage{graphicx}
\usepackage{amsfonts}
\usepackage{amssymb}
\usepackage{epsfig}
\usepackage{color}
\usepackage{psfrag}
\usepackage{epstopdf}

\newcommand{\EQ}{\begin{equation}}
\newcommand{\EN}{\end{equation}}
\newcommand{\be}{\begin{equation}}
\newcommand{\ee}{\end{equation}}
\newcommand{\bea}{\begin{eqnarray}}
\newcommand{\eea}{\end{eqnarray}}

\newcommand{\rd}{{\rm d}}

\begin{document} \setcounter{page}{0}
\topmargin 0pt
\oddsidemargin 5mm
\renewcommand{\thefootnote}{\arabic{footnote}}
\newpage
\setcounter{page}{0}
\topmargin 0pt
\oddsidemargin 5mm
\renewcommand{\thefootnote}{\arabic{footnote}}
\newpage
\begin{titlepage}
\begin{flushright}
\end{flushright}
\vspace{0.5cm}
\begin{center}
{\large {\bf Fields, particles and universality in two dimensions}}\\
\vspace{1.8cm}
{\large Gesualdo Delfino}\\
\vspace{0.5cm}
{\em SISSA -- Via Bonomea 265, 34136 Trieste, Italy}\\
{\em INFN sezione di Trieste}\\
{\em E-mail: delfino@sissa.it}\\
\end{center}
\vspace{1.2cm}

\renewcommand{\thefootnote}{\arabic{footnote}}
\setcounter{footnote}{0}

\begin{abstract}
\noindent
We discuss the use of field theory for the exact determination of universal properties in two-dimensional statistical mechanics. After a compact derivation of critical exponents of main universality classes, we turn to the off-critical case, considering systems both on the whole plane and in presence of boundaries. The topics we discuss include magnetism, percolation, phase separation, interfaces, wetting.
\end{abstract}
\end{titlepage}

\newpage

\tableofcontents

\section{Introduction}
\label{introduction}
Statistical systems close to a second order phase transition point exhibit properties which do not depend on details of the microscopic interaction but only on global features such as internal symmetries and space dimensionality. These properties are called universal, and systems sharing them are said to belong to the same universality class. Renormalization theory has progressively clarified how the divergence of the correlation length as the critical point is approached allows for the emergence of universality, and how field theory actually is the theory of universality classes.

While these ideas apply in any dimension $d\geq 2$, to the point that $d$ serves as an expansion parameter in the Wilson-Fisher approach to renormalization, different dimensionalities exhibit specific features which are intrinsically non-perturbative and determine distinctive qualitative properties of universal behavior. The two-dimensional case, in particular, is characterized by features such as absence of spontaneous breaking of continuous symmetries, fermionization (or, conversely, bosonization), correspondence between interfaces and particle trajectories, infinite dimensional conformal symmetry at criticality. The last property is in turn at the origin of a further, remarkable peculiarity of the two-dimensional case, i.e. the existence of exact solutions, both at criticality and near criticality, and for all universality classes.

The aim of this article is to illustrate how field theory leads to the exact description of universality classes of classical equilibrium statistical mechanics in two dimensions. The presentation focuses on basic ideas and examples, with a minimal amount of technical aspects.  It relies on several recent results, and is original for the spectrum of the topics discussed, ranging from percolation to wetting, and the way they are cast into a unified theoretical framework.

The discussion is structured according to the following path. In the next section we recall generalities about universality, field theory and its particle description. In section~\ref{conformal} we give a concise overview of two-dimensional conformal field theory suitable for our subsequent discussion; in particular, we do not focus on minimal models, leaving room for cases, like percolation, in which the bulk correlation functions of the order parameter do not satisfy known differential equations. In section~\ref{scale} we show how the critical indices of main universality classes, including percolation and self-avoiding walks, can be identified directly in field theory exploiting the insight coming from the particle description. In section~\ref{away} we recall how, under conditions which turn out to account for the main interesting cases, exact solvability extends to the off-critical regime, and how solutions originally obtained in the scattering framework lead to the computation of universal quantities such as combinations of critical amplitudes. The extension of the formalism to crossover phenomena is illustrated in section~\ref{crossover}. In section~\ref{phase_separation} we turn to systems with boundaries, showing how general low-energy properties of two-dimensional field theory yield the exact characterization of phase separation and of the interfacial region, an analysis extended in section~\ref{interfacesatboundaries} to describe the interaction of an interface with the boundary and the effect of boundary geometry. In section~\ref{nearcritical} we  show how the particle formalism provides exact asymptotic results for percolation on the rectangle away from criticality, where the methods of boundary conformal field theory do not apply. The last section is devoted to few final remarks.

\section{General notions}
\label{general}
\subsection{Universality}
\label{universality}
In the framework of classical equilibrium statistical mechanics \cite{LL} a system is specified by the Hamiltonian ${\cal H}$, whose value is determined by the configuration of the system. The expectation value of an observable ${\cal O}$ is the statistical average over configurations
\EQ
\langle{\cal O}\rangle=\frac{1}{Z}\sum_\textrm{configurations}{\cal O}\,e^{-{\cal H}/T},
\EN
where $T\geq 0$ is the temperature\footnote{We adopt units in which $k_B=1$.} and the normalization factor (partition function)
\EQ
Z=\sum_\textrm{configurations}e^{-{\cal H}/T}
\EN
ensures that the average of the identity is 1. We are interested in systems which in the limit of infinitely many degrees of freedom undergo a phase transition for some critical value $T_c$ of the temperature. In more than one infinite dimension, this is signaled by an order observable whose expectation value (order parameter) vanishes above $T_c$ and is a function of $T$ below $T_c$. The transition is said to be of the first order if the order parameter has a discontinuity at $T_c$, and  of the second order (or, more generally, continuous) otherwise.

Phase transitions are normally associated to {\em spontaneous} symmetry breaking. A group $G$, mapping configurations into configurations, is a symmetry of the system if ${\cal H}$ is left invariant by the action of $G$. Configurations of minimal energy (ground states) are mapped into each other by $G$, and below $T_c$ a $G$-invariant system chooses the phase dominated by a specific ground state as $T\to 0$. 

As $T\to T_c$, if the transition is continuous, universal critical properties emerge, i.e. quantitative properties common to systems having the same dimensionality and symmetry group. Such systems are said to belong to the same {\em universality class}. The Ising model is the basic representative of the simplest universality class, that associated to the spontaneous breaking of the group $G=Z_2$. The model is characterized by the Hamiltonian 
\EQ
{\cal H}_\textrm{Ising}=-J\sum_{\langle i,j\rangle}\sigma_i\sigma_j\,,\hspace{1cm}\sigma_i=\pm 1\,,
\label{ising}
\EN
where a ``spin'' variable $\sigma_i$ is assigned to the $i$-th site of a regular lattice and the sum is restricted to the pairs of nearest neighboring sites. Unless otherwise stated, we refer to the ferromagnetic case $J>0$, for which the model possesses two ground states (spins all plus or all minus); the $Z_2$ transformation corresponds to the reversal of all spins, and leaves (\ref{ising}) invariant. The order parameter $\langle\sigma_i\rangle$ is proportional to $(T_c-T)^\beta$ as $T\to T_c^-$, with a {\it critical exponent} $\beta$ which is universal. This can be contrasted with the value of $T_c$, which changes, for example, when the structure of the lattice changes, or next to nearest neighbors interactions are included. The spin-spin correlation function $\langle\sigma_i\sigma_j\rangle$ decays as $e^{-|i-j|/\xi}$ as the distance $|i-j|$ between the two spins goes to infinity; such a decay provides the definition of the correlation length $\xi$ that we use throughout this article. Continuous phase transitions are characterized by the fact that the correlation length diverges as $\xi\sim\xi_0^{\pm}|T-T_c|^{-\nu}$ when $T\to T_c^\pm$, so that exponential decay of correlation functions is replaced at $T_c$ by power law decay. It is the divergence of the correlation length which makes microscopic details irrelevant close to criticality and leads to the emergence of universal behavior. The susceptibility $\chi=1/N\sum_j\langle\sigma_i\sigma_j\rangle$ ($N\to\infty$ is the number of sites) diverges as 
\EQ
\chi\sim\Gamma_\pm|T-T_c|^{-\gamma}\,,\hspace{1cm}T\to T_c^\pm\,.
\label{chi}
\EN
While the exponents $\nu$ and $\gamma$ are universal, the critical amplitudes $\xi_0^\pm$ and $\Gamma_\pm$ are not; $\xi_0^+/\xi_0^-$ and $\Gamma_+/\Gamma_-$, however, are universal, since non-universal "metric factors" cancel in the ratio. 

The $q$-state Potts model, defined by the lattice Hamiltonian \cite{Potts,Wu} 
\EQ
{\cal H}_\textrm{Potts}=-J\sum_{\langle i,j\rangle}\delta_{s_i,s_j}\,,\hspace{1cm}s_i=1,2,\ldots,q\,,
\label{potts}
\EN
generalizes the Ising model to the case in which the site variable takes $q$ values, to which we will often refer as $q$ ``colors''. The Hamiltonian is invariant under the action of the group $G=S_q$ of permutations of the colors, and the model possesses $q$ ferromagnetic ground states in which all sites have the same color. The expectation value of the spin variable
\EQ
\sigma_{i,\alpha}\equiv\delta_{s_i,\alpha}-\frac{1}{q}\,,\hspace{1cm}\alpha=1,\ldots,q\,,
\label{potts_spin}
\EN
vanishes as long as the symmetry is unbroken and provides the order parameter for the transition. A particularly interesting feature of the $q$-state Potts model is that the partition function admits the graph expansion \cite{FK}
\EQ
Z_\textrm{Potts}=\sum_{\{s_i\}}e^{-{\cal H}_\textrm{Potts}/T}\propto\sum_G p^{N_b}(1-p)^{\bar{N}_b}q^{N_c}\,,
\label{FK}
\EN
where $G$ is a graph obtained drawing bonds on the edges of the lattice, $N_b$ is the number of bonds in $G$, $\bar{N}_b$ the number of edges without a bond, and $p=1-e^{-J/T}$; a set of connected bonds is called a cluster, and $N_c$ is the number of clusters in $G$, with the convention that a site not touched by any bond is also counted as a cluster. The graph expansion makes sense of the Potts partition function for continuous values of $q$. In particular, for $q\to 1$ the weight $p^{N_b}(1-p)^{\bar{N}_b}$ of a configuration coincides with that of the bond percolation problem \cite{SA,Grimmett}, in which edges are occupied with probability $p$ and empty with probability $1-p$. Percolation is characterized by the existence of a critical threshold $p_c$ such that for $p>p_c$ the probability of finding an infinite cluster is larger than zero. Equation (\ref{FK}) relates such a ``geometric'' phase transition to the thermodynamic phase transition of the $q\to 1$ Potts model.

The basic model exhibiting a continuous symmetry is the vector model defined by the Hamiltonian
\EQ
{\cal H}_\textrm{vector}=-J\sum_{\langle i,j\rangle}{\bf s}_i\cdot{\bf s}_j\,,
\label{vector}
\EN
where the spin variable ${\bf s}_i$ is a $n$-component unit vector. The symmetry group is $G=O(n)$ and the case $n=2$ is usually called XY model, while the Ising model is recovered for $n=1$. The partition function of the vector model also admits a graph expansion in which $n$ enters as a parameter that can be taken continuous. More precisely, the mapping involves a Hamiltonian slightly different from (\ref{vector}) (see e.g. \cite{Cardy_book}), but belonging to the same universality class. The graphs are configurations of closed paths (loops), in which each loop contributes a factor of $n$ to the weight. The limit $n\to 0$ is dominated by the configurations with a single loop and describes the self-avoiding polymer problem \cite{deGennes_walks}; we will see in section~\ref{sq} how self-avoidance emerges as a property of the universality class.

\subsection{Fields}
\label{fields}
The independence of universal properties on microscopic details suggests that universality is most naturally captured within a continuous description. Field theory provides such a continuous framework for the study of universality classes (see e.g. \cite{Cardy_book} and references therein). For a $d$-dimensional system we denote by $x=(x_1,\ldots,x_d)$ a point in the Euclidean space $R^d$, and replace site-dependent lattice observables by $x$-dependent {\em fields}. In particular, the lattice spin variable is replaced by a spin (or order) field, $\sigma(x)\in R$ for the Ising model. Similarly, to the energy density $\sum_j'\sigma_i\sigma_j$ entering (\ref{ising}) (the prime indicates summation over nearest neighbors of the site $i$) we associate an energy density field $\varepsilon(x)$. The Euclidean action (or reduced Hamiltonian) ${\cal A}$ is a functional of the fields which specifies the theory and determines the weight of a field configuration as $e^{-{\cal A}}/Z$, where we keep the notation $Z$ for the partition function $\sum_\textrm{field\,configs}e^{-{\cal A}}$ in the continuum. 

The action ${\cal A}$ is invariant under the action of a symmetry group $G$ of the system. A further characterization can be achieved starting from critical point actions \cite{WK}. The divergence of the correlation length and the power law decay of correlation functions at a second order phase transition point amount to {\em scale invariance} of the critical point field theory. Such a theory contains no dimensional couplings and is left invariant by scale transformations $x\to\alpha\,x$; it is usually called a ``fixed point'' theory and in this section we will denote fixed point quantities by a subscript $FP$. Scale invariance leads to the two-point function
\EQ
\langle\Phi(x_1)\Phi(x_2)\rangle_{FP}=\frac{\textrm{constant}}{|x_1-x_2|^{-2X_\Phi}}
\label{power}
\EN
for a scaling field $\Phi(x)$; this equation serves as a definition for the {\em scaling dimension} $X_\Phi$ of the field at the given fixed point. The off-critical theory can now be associated to the action
\EQ
{\cal A}={\cal A}_{FP}+\tau\int d^dx\,\varepsilon(x)\,,
\label{scaling}
\EN
where $\tau$ measures the deviation from $T_c$. Since the action is dimensionless, $\tau$ has the dimension of an inverse length (a mass in our units) to the power $d-X_\varepsilon$ and provides the dimensional coupling which breaks scale invariance. On dimensional grounds we have $\xi\propto|\tau|^{-1/(d-X_\varepsilon)}$, and then the critical exponent $\nu=1/(d-X_\varepsilon)$.

A $G$-invariant field theory contains infinitely many fields with growing scaling dimension which transform in the same way under the action of the group. In writing (\ref{scaling}) we omitted infinitely many $G$-invariant fields with scaling dimensions larger than $d$. Their conjugated couplings have the dimension of a length to positive powers and become negligible when the system is observed over distances much larger than lattice spacing, i.e. in the limit in which universal properties emerge. In this sense such fields are called ``irrelevant''. The action in which irrelevant fields are omitted is called the scaling action and characterizes the universality class. The scaling action may contain more than one $G$-invariant field with scaling dimension smaller than $d$ (such fields are called ``relevant''). This corresponds to theories in which more than one parameter needs to be tuned to achieve scale invariance; these theories describe ``multicritical'' behavior. A theory may contain fields with scaling dimension equal to $d$ (``marginal'' fields). Marginality may be spoiled by logarithmic corrections produced by off-critical interactions. If this is not the case, the addition to the fixed point action of such a ``truly marginal'' field does not break scale invariance and generates a line of fixed points. The recognition that the properties of the system depend on the scale at which it is observed and the characterization in terms of fixed points and relevant and irrelevant fields form the basis of the {\it renormalization group} idea \cite{WK,Cardy_book}.

We saw how the correlation length critical exponent is determined by $X_\varepsilon$. On the other hand, scaling dimensional analysis gives $\langle\sigma(x)\rangle\propto\xi^{-X_\sigma}$, and then $\beta=\nu X_\sigma$. Similarly, the susceptibility $\chi=\int d^dx\langle\sigma(x)\sigma(0)\rangle$ scales as $\xi^{d-2X_\sigma}$, so that $\gamma=\nu(d-2X_\sigma)$. Hence the theory accounts for the universality of the critical exponents and for the relations among them, and reduces their determination to that of the scaling dimensions of the energy density and order fields. Also combinations of critical amplitudes like $\xi_0^+/\xi_0^-$ and $\Gamma_+/\Gamma_-$, being dimensionless, are universal. It also follows that, while critical exponents are determined by the fixed point theory, critical amplitudes need to be computed within the off-critical theory (\ref{scaling}).

\subsection{Particles}
\label{particles}
We are considering the ordinary case of systems that are isotropic in the continuum limit, so that the scaling action (\ref{scaling}) is invariant under rotations. Such an Euclidean field theory defines, when one of the coordinates is made purely imaginary ($x_d=it$), a quantum field theory with spatial coordinates $x_1,\ldots,x_{d-1}$ and time coordinate $t$, in which relativistic invariance replaces $d$-dimensional rotational invariance. The quantum theory has the same field content as the Euclidean one and correlation functions in the two cases are related by analytic continuation from real to imaginary time. The quantum theory also admits a description in terms of relativistic particles corresponding to the excitations above a minimum energy (vacuum) state; spontaneous symmetry breaking corresponds to the presence of degenerate vacua mapped into one another by the symmetry \cite{Weinberg}.

\begin{figure}
\begin{center}
\includegraphics[width=3cm]{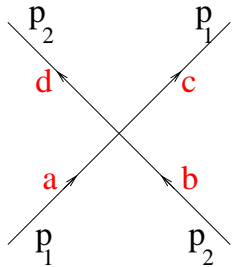}
\caption{Elastic two-particle process in 1+1 dimensions. Time runs upwards.}
\label{fpu_elastic}
\end{center} 
\end{figure}

The particle description of a quantum field theory is encoded by the $S$-matrix \cite{ELOP}, whose matrix elements are the probability amplitudes that a particle state at time $t=-\infty$ evolves into another state at time $t=+\infty$. In general, relativistic scattering conserves total energy and momentum, but not the number of particles. We call ``elastic'' a scattering process in which the number of particles and their masses are preserved. Specializing to the case $d=2$, to which we turn from now on, we depict in Fig.~\ref{fpu_elastic} a two-particle elastic process. We consider the case in which the particles have the same mass $m$. The energy $e$ and the momentum $p$ of a particle are related by the relativistic dispersion relation $e=\sqrt{p^2+m^2}$. For the process we are considering, conservation of energy and momentum imply that the momenta $p_1$ and $p_2$ are individually conserved by the scattering. If the particles carry charges, the total charge may be redistributed in the scattering, a possibility that we take into account through the particle species labels $a,b,c,d$.

The scattering amplitude of Fig.~\ref{fpu_elastic} is relativistically invariant and depends on the single relativistic invariant quantity that can be built out of the two energy-momenta; we take this invariant in the form of the square of the center of mass energy
\EQ
s=(e_1+e_2)^2-(p_1+p_2)^2\,.
\label{e2}
\EN
The amplitude, that we denote as ${\cal S}_{ab}^{cd}(s)$, satisfies a number of properties \cite{ELOP} that we now state. First of all the amplitude is an element of the {\em unitary} $S$-matrix. When continued to complex values of the variable $s$, it is an analytic function whose singularities have a physical implication. In particular, as a consequence of unitarity, the minimal energy values (thresholds) needed for the production of a new final state correspond to branch points of the amplitude; these are located at $s=(km)^2$ for the production of $k>1$ particles of mass $m$. A simple pole at $s=\tilde{m}^2$ corresponds instead to a particle of mass $\tilde{m}$ appearing as a stable bound state of the two particles in the initial (or final) state of Fig.~\ref{fpu_elastic}.

\begin{figure}
\begin{center}
\includegraphics[width=12cm]{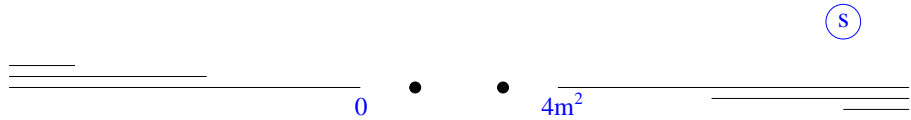}
\caption{Analytic structure of the scattering amplitude in the complex $s$-plane. Unitarity (right) and crossing (left) cuts are shown together with two poles.}
\label{fpu_s_plane}
\end{center} 
\end{figure}

The unitarity cuts originating from the branch points we just discussed are taken along the positive real axis in the complex $s$-plane as in Fig.~\ref{fpu_s_plane}, and the physical value of the amplitude corresponds to the limit towards the real axis from above (${\cal S}_{ab}^{cd}(s+i\epsilon)$ with $s\in R$, $\epsilon\to 0^+$). Such a path lies on the first (called ``physical'') sheet of the Riemann surface associated to the cut $s$-plane; additional sheets are accessed sliding into the cuts.

A further fundamental property of relativistic scattering is {\it crossing symmetry}. It states that the amplitude for the direct channel process (Fig.~\ref{fpu_elastic} with time running upwards) is related by analytic continuation to the amplitude for the crossed channel process (Fig.~\ref{fpu_elastic} with time running from left to right). In the passage to the crossed channel, the particles $b$ and $d$, which have the arrow pointing in the `wrong' direction, are switched into antiparticles $\bar{b}$ and $\bar{d}$, and have their energy and momentum reversed ($e_2,p_2\to-e_2,-p_2$, leading to $s\to 4m^2-s$). The crossing relation then reads
\EQ
{\cal S}_{ab}^{cd}(s+i\epsilon)={\cal S}_{\bar{d}a}^{\bar{b}c}(4m^2-s-i\epsilon)\,,
\label{crossing0}
\EN
for $s$ real. A consequence of (\ref{crossing0}) is that an amplitude inherits from the crossed channel branch cuts running along the negative real axis, as well as crossing images of bound state poles (Fig.~\ref{fpu_s_plane}). 

We finally mention the property of {\it real analyticity}, which states that the values of the amplitude on the upper and lower edge of a cut are related by complex conjugation; this implies in particular that the amplitude is real along the uncut portion of the real axis. It also allows one to write the unitarity condition as
\EQ
\sum_{e,f}{\cal S}_{ab}^{ef}(s+i\epsilon){\cal S}_{ef}^{cd}(s-i\epsilon)=\delta_{ac}\delta_{bd}\,,\hspace{1cm}4m^2<s<s_1\,,
\label{unitarity0}
\EN
for physical values of $s$ below the first inelastic threshold $s_1$.

Within the particle framework, fields correspond to operators (we denote them by the same symbol) which acting on a particle state produce a superposition of particle states. The Hamiltonian and momentum operators $H$ and $P$ of the $(1+1)$-dimensional quantum system act as generators of time and space translations and, in Euclidean coordinates $x_1$ and $x_2=it$, we have
\EQ
\Phi(x)=e^{-iPx_1+Hx_2}\Phi(0)e^{iPx_1-Hx_2}\,.
\label{shift}
\EN
The field/operator is characterized by the {\it form factors} 
\EQ
{\cal F}^\Phi_n(p_1,\ldots,p_n)=\langle 0|\Phi(0)|p_1,\ldots p_n\rangle\,,
\label{ff}
\EN
where we are considering a single particle species in order to simplify the notation, $|0\rangle$ denotes the vacuum state (i.e. the state without particles) and $|p_1,\ldots p_n\rangle$ are asymptotic particle states, i.e. corresponding to $t=-\infty$ (incoming states) or $t=+\infty$ (outgoing states) in the scattering picture; in these states the particles are widely separated in space and can be treated as non-interacting. The asymptotic states are eigenstates of $H$ and $P$ with eigenvalues $\sum_{i=1}^ne_i$ and $\sum_{i=1}^np_i$, respectively. The two-point function of a scalar field decomposes over the basis of asymptotic states as
\bea
\langle\Phi(x)\Phi(0)\rangle &=& \langle 0|\Phi(x)\Phi(0)|0\rangle\nonumber\\
&=&\sum_{n=0}^\infty\frac{1}{n!}\int\frac{dp_1}{2\pi e_1}\cdots\frac{dp_n}{2\pi e_n}|{\cal F}^\Phi_n(p_1,\ldots,p_n)|^2e^{-|x|\sum_{i=1}^ne_i}\,,
\label{spectral}
\eea
where we take advantage of rotational invariance to set $x_1=0$ and adopt the state normalization $\langle p_1|p_2\rangle=2\pi e_1\delta(p_1-p_2)$. For $|x|\to\infty$ the correlator decays exponentially, showing that the mass $m$ is inversely proportional to the correlation length. If ${\cal F}^\Phi_1(p)\neq 0$ we say that $\Phi(x)$ creates the particle.

\section{Critical points}
\label{conformal}
\subsection{Conformal symmetry}
\label{basic}
In writing the two-point function (\ref{power}) we assumed that the field $\Phi$ is scalar, i.e. invariant under rotations. More generally, in two dimensions we will consider scaling fields $\Phi(x)$ with scaling dimension $X_\Phi$ which transform as $\Phi(0)\to e^{-is_\Phi\alpha}\Phi(0)$ under a rotation by an angle $\alpha$ centered in the origin; $s_\Phi$ is the Euclidean spin of the field. It is an important property of field theory that products of local fields can be expanded onto an infinite-dimensional basis of local fields. In a scale invariant two-dimensional theory the need to preserve dimensional and rotational properties gives to such an operator product expansion (OPE) the form
\bea
\Phi_i(x)\Phi_j(0)&=&\sum_k C_{ij}^{k}\,(z\bar{z})^{(X_k-X_i-X_j)/2}(z\bar{z}^{-1})^{(s_k-s_i-s_j)/2}\,\Phi_k(0)\nonumber\\
&=&\sum_k C_{ij}^{k}\,z^{\Delta_{k}-\Delta_{i}-\Delta_{j}}\bar{z}^{\bar{\Delta}_{k}-\bar{\Delta}_{i}-\bar{\Delta}_{j}}\,\Phi_k(0)\,,
\label{ope}
\eea
where we introduced the complex coordinates $z=x_1+ix_2$ and $\bar{z}=x_1-ix_2$, which rotate as  $z\to e^{i\alpha}z$, $\bar{z}\to e^{-i\alpha}\bar{z}$, as well as the left and right dimensions $\Delta_\Phi$, $\bar{\Delta}_\Phi$ such that $X_\Phi=\Delta_\Phi+\bar{\Delta}_\Phi$ and  $s_\Phi=\Delta_\Phi-\bar{\Delta}_\Phi$ ($X_{\Phi_i}=X_i$, ...); the coefficients $C_{ij}^k$ are called structure constants. Two fields $\Phi_i$ and $\Phi_j$ are said to be mutually local if the correlation functions $\langle\cdots\Phi_i(x)\Phi_j(0)\cdots\rangle$ are single valued under the continuation $z\to e^{2i\pi}z$, $\bar{z}\to e^{-2i\pi}\bar{z}$. This is the case if
\EQ
s_k-s_i-s_j\in Z
\label{gamma}
\EN
for all $\Phi_k$ contributing to the r.h.s. of (\ref{ope}).

A field theory contains a symmetric and conserved energy-momentum tensor\footnote{In real time, energy and momentum are the integrals over $x_1$ of $T_{tt}$ and $T_{tx_1}$, respectively.} $T_{\mu\nu}(x)$ with scaling dimension $X_{T_{\mu\nu}}=d$. In two dimensions the three independent components in the basis of complex coordinates are $T_{zz}\equiv T$, $T_{\bar{z}\bar{z}}\equiv\bar{T}$ and $T_{z\bar{z}}=T_{\bar{z}z}\equiv-\Theta$, with dimensions $(\Delta,\bar{\Delta})$ equal to $(2,0)$, $(0,2)$ and $(1,1)$, respectively; the conservation equations read 
\EQ
\bar{\partial}T=\partial\Theta\,,\hspace{1cm}\partial\bar{T}=\bar{\partial}\Theta\,,
\label{Tconservation}
\EN
 where $\partial\equiv\partial_z$ and $\bar{\partial}\equiv\partial_{\bar{z}}$. In a scale invariant theory, $\bar{\Delta}_T=\Delta_{\bar{T}}=0$ means $T=T(z)$, $\bar{T}=\bar{T}(\bar{z})$, i.e. $\bar{\partial}T=\partial\bar{T}=0$, so that the trace $T_\mu^\mu=-4\Theta$ is a constant; $X_\Theta\neq 0$ then implies that the energy-momentum tensor is traceless in presence of scale invariance, a property that actually holds also in higher dimensions (see e.g. \cite{Cardy_book}).

Within the theories we are interested in for physical applications all fields are mutually local with the energy-momentum tensor. We then have the OPE
\EQ
T(z)\Phi(0)=\sum_{n=-\infty}^{\infty}z^{-2-n}L_n\Phi(0)\,,
\label{TPhi}
\EN
where the fields $L_n\Phi$ have dimensions $(\Delta,\bar{\Delta})=(\Delta_\Phi-n,\bar{\Delta}_\Phi)$. The $L_n$'s then act as shift operators in the space of fields graded by $\Delta$ (similarly for the $\bar{L}_n$ one obtains from $\bar{T}$). There will be fields $\phi$ with lowest dimension ({\it primary fields}) characterized by the property $L_n\phi=0$ for any positive $n$; the fields $L_n\phi$ with $n<0$ are examples of the so-called {\it descendants} of the primary $\phi$; $T=L_{-2}I$ is a descendant of the identity. A primary $\phi$ and its descendants form an operator family that we denote by $[\phi]$, and the collection of the operator families forms the space of fields.

An additional property of scale invariant field theories is that they are actually invariant under the more general group of conformal transformations, i.e. the transformations which allow for a point dependent multiplicative change of the infinitesimal distance element (see e.g. \cite{Cardy_book}); in this sense they are called conformal field theories. In two dimensions infinitesimal conformal transformations correspond to changes $\delta z=f(z)$, $\delta\bar{z}=\bar{f}(\bar{z})$, where $f$ (resp. $\bar{f}$) is any analytic function of $z$ (resp. $\bar{z}$), in such a way that the group of conformal transformations in two dimensions has the essential peculiarity of being infinite-dimensional. This turns out to imply \cite{BPZ,DfMS} that the operators $L_n$ entering (\ref{TPhi}) satisfy the Virasoro algebra
\EQ
[L_n,L_m]=(n-m)L_{n+m}+\frac{c}{12}(n^3-n)\delta_{n,-m}\,,
\label{Virasoro}
\EN
where $c$ is a fundamental parameter of the theory known as {\it central charge}. One has $L_0\Phi=\Delta_\Phi\Phi$, $L_{-1}\Phi=\partial\Phi$ and 
\EQ
T(z)T(0)=\frac{c}{2z^4}+\frac{2}{z^2}T(0)+\frac{1}{z}\partial T(0)+\ldots\,.
\label{TT}
\EN
The central charge of theories which are the sum of sub-systems non-interacting with each other is the sum of the central charges of the sub-systems and, more generally, it can be shown that $c$ grows with the number of degrees of freedom of the critical point \cite{Z_cth}. Hence, for example, Ising, three-state Potts and XY universality classes have to correspond to increasing values of the central charge.

The algebra (\ref{Virasoro}) provides a powerful tool for the study of two-dimensional conformal field theories, since the operator families correspond to lowest weight representations of the algebra. The dimension $\Delta_\phi$ (which is also called ``conformal'' dimension) of the primary $\phi$ is the lowest weight, and the subspace of descendants with dimension $\Delta_\phi+l$ is spanned by the fields $L_{-j_1}\ldots L_{-j_J}\phi$ with $0<j_1\leq\ldots\leq j_J$ and $\sum_{n=1}^Jj_n=l$; hence, generically, the dimension of such a subspace is given by the number $p(l)$ of partitions of $l$ (the {\it level}) into positive integers. A very important role in the theory is played by the reducible representations of the algebra (\ref{Virasoro}), i.e. the representations $[\phi]$ which contain another representation $[\phi_0]$ whose primary $\phi_0$ is a descendant of $\phi$ at some level $l_0$. The irreducible representation obtained factoring out $[\phi_0]$ is said to be a representation {\it degenerate} at level $l_0$, and $\phi$ a degenerate primary. Clearly, in such a degenerate representation the number of independent fields at level $l\geq l_0$ is smaller than $p(l)$. Factoring out $[\phi_0]$ amounts to setting to zero $\phi_0$ and its descendants, a condition which can be shown to lead to linear partial differential equations for multi-point correlation functions containing a degenerate primary \cite{BPZ}.

The basic universality classes of two-dimensional critical behavior correspond to conformal theories with central charge $c\leq 1$. As pointed out in \cite{DF}, these can be described starting from the theory of a free scalar field with action
\EQ
{\cal A}_0=\frac{1}{4\pi}\int d^2x\,(\nabla\varphi)^2\,.
\label{free}
\EN
The dimensionless character of $\varphi(x)$ leads to the logarithmic correlator
\EQ
\langle\varphi(x)\varphi(0)\rangle=-\ln|x|=-\frac{1}{2}(\ln z+\ln\bar{z})\,,
\label{log}
\EN
which is consistent with the equation of motion $\partial\bar{\partial}\varphi=0$ and the decomposition 
\EQ
\varphi(x)=\chi(z)+\bar{\chi}(\bar{z})\,.
\EN
The components of the energy-momentum tensor can be written in the form
\EQ
T(z)=-(\partial\varphi)^2+iQ\,\partial^2\varphi\,,\hspace{1cm}
\bar{T}(\bar{z})=-(\bar{\partial}\varphi)^2+iQ\,\bar{\partial}^2\varphi\,;
\label{T}
\EN
the presence of the terms containing the parameter $Q$ is allowed by the fact that, due to the equation of motion, they do not affect conservation and are total derivatives with respect to $x_1$, so that they do not contribute to energy and momentum. Using (\ref{log}) and Wick theorem, we can compute $T(z)T(w)$ and, comparing with (\ref{TT}), obtain the central charge
\EQ
c=1-6Q^2\,.
\label{c}
\EN
Instead of $\chi$, which has $\langle\chi(z)\chi(0)\rangle\propto\ln z$, proper scaling primary fields of the theory are the exponentials
\EQ
V_{p}(z)=e^{2ip\chi(z)}\,,
\label{V}
\EN
 whose dimension $\Delta_p$ is obtained considering $T(z)V_p(w)$ and isolating the term $\Delta_pV_p(w)(z-w)^{-2}$ (the term with $n=0$ in (\ref{TPhi})); this yields
\EQ
\Delta_p=p(p-Q)\,.
\label{deltap}
\EN
Of course we also have $\bar{V}_{\bar{p}}(\bar{z})=e^{2i\bar{p}\bar{\chi}(\bar{z})}$, in such a way that the generic primary $V_p\bar{V}_{\bar{p}}$ has dimensions $(\Delta_p,\Delta_{\bar{p}})$. 

We have already seen that at criticality the requirement of mutual locality of the fields with the energy-momentum tensor is implemented by (\ref{TPhi}). When moving to the off-critical action (\ref{scaling}), the energy-momentum tensor acquires a non-zero trace proportional to the field responsible for the breaking of scale invariance: $\Theta(x)\propto\tau\varepsilon(x)$. We now examine, first for $c=1$ and then for $c<1$, some implications of the choice of $\varepsilon$ and of the requirement of locality with respect to $\varepsilon$.

\subsection{$c=1$}
\label{c1}
Central charge equal 1 corresponds to the case $Q=0$, namely to the usual Gaussian theory with OPE
\EQ
V_{p_1}\cdot V_{p_2}=\left[V_{p_1+p_2}\right]\,,
\label{gaussian}
\EN
where, with respect to (\ref{ope}), we omit for brevity coordinate dependence and structure constants; the brackets indicate that, besides the primary field, the r.h.s. contains its descendants. Using (\ref{gamma}), (\ref{deltap}) and (\ref{gaussian}) we obtain that two fields $V_{p_1}\bar{V}_{\bar{p}_1}$ and $V_{p_2}\bar{V}_{\bar{p}_2}$ are mutually local if 
\EQ
2(p_1p_2-\bar{p}_1\bar{p}_2)\in Z\,.
\label{gaussian_locality}
\EN
A generic choice for a real and scalar energy density field in this theory is $\varepsilon=V_b\bar{V}_b+V_{-b}\bar{V}_{-b}\propto\cos 2b\varphi$ with $\Delta_\varepsilon=b^2$; a field $V_p\bar{V}_{\bar{p}}$ is local with respect to $\varepsilon$ if $\pm 2b(p-\bar{p})$ is an integer, i.e. if 
\EQ
p-\bar{p}=\frac{m}{2b}\,,\hspace{1cm}m\in Z\,.
\label{quantization}
\EN
Taking $m=1$, we can build the complex (Dirac) fermion 
\EQ
\Psi=(\psi,\bar{\psi})=\left(V_{\frac{1}{4b}+\frac{b}{2}}\bar{V}_{-\frac{1}{4b}+\frac{b}{2}},V_{\frac{1}{4b}-\frac{b}{2}}\bar{V}_{-\frac{1}{4b}-\frac{b}{2}}\right)\,,
\label{dirac}
\EN
with spin $p^2-\bar{p}^2$ equal $1/2$ for $\psi$ and $-1/2$ for $\bar{\psi}$; the decomposition $\Psi=\Psi_1+i\Psi_2$ defines two real (Majorana) fermions $\Psi_i=(\psi_i,\bar{\psi}_i)$. For $b^2=1/2$ we have $\bar{\partial}\psi=\partial{\bar \psi}=0$, which are free fermionic equations of motion; hence, for $b^2=1/2$, the theory (\ref{free}) admits a representation in terms of free fermions. For $b^2\neq 1/2$, instead, the fermions are coupled by the four-fermion term, which can be shown to be truly marginal, so that the bosonic action (\ref{free}) with energy density field $\cos 2b\varphi$ fermionizes into \cite{Coleman,Mandelstam,ZZ_cft}
\EQ
{\cal A}_0=\int d^2x\,\left[\sum_{i=1,2}(\psi_i\bar{\partial}\psi_i+\bar{\psi}_i\partial\bar{\psi_i})+g(b^2)\psi_1\bar{\psi_1}\psi_2\bar{\psi_2}\right]\,,
\label{Thirring}
\EN
with $g(1/2)=0$; $\cos2b\varphi$ fermionizes into the mass term $\psi_1\bar{\psi_1}+\psi_2\bar{\psi_2}$. 

Since the real fermionic fields satisfy $\psi_i^2=\bar{\psi}_i^2=0$, so that $\psi_1\bar{\psi_1}\psi_2\bar{\psi_2}\propto(\sum_i\psi_i\bar{\psi}_i)^2$, the action (\ref{Thirring}) is invariant under the $O(2)$ rotations of the vector $(\psi_1,\psi_2)$; these become $U(1)$ transformations for the complex fermion, and the integer $m$ in (\ref{quantization}) plays the role of $U(1)$ charge. Then the components of the order field associated to the $U(1)\sim O(2)$ symmetry are selected picking up the scalar fields with $m=\pm 1$, i.e. $\sigma_\pm=V_{\pm 1/4b}\bar{V}_{\mp 1/4b}$ with $\Delta_{\sigma_\pm}=1/16b^2$. If, by construction, $\cos 2b\varphi$ is the most relevant $O(2)$-invariant field, generic fields entering the most general action with $O(2)$ symmetry must have $m=0$ and must be local with respect to the order field components $\sigma_\pm$; (\ref{gaussian_locality}) then identifies these fields with $e^{2ikb\varphi}$, with $k$ integer and conformal dimension $k^2b^2$. We then see that, ignoring the identity ($k=0$), all these fields are irrelevant for $b^2>1$; $\cos 2b\varphi$ is the only invariant relevant field for $1/4<b^2<1$ and leads to an off-critical action (known as sine-Gordon model) which, as we will see later, is purely massive for both signs of the conjugated coupling. This leaves no room for a phase with spontaneously broken $O(2)$ symmetry, which requires a relevant invariant field and massless (Goldstone) excitations associated to a continuum of degenerate vacua. Indeed, the two-dimensional $XY$ model only exhibits a transition between a high-temperature phase with exponential decay of correlations and a low-temperature phase with algebraic decay, while the order parameter vanishes at all temperatures. This is known as Berezinsky-Kosterlitz-Thouless (BKT) transition \cite{KT} and is described by the Gaussian model with $b^2(T)$ a (non-universal) decreasing function of the temperature; $b^2=1$ at the BKT transition temperature $T_{BKT}$, so that the low-temperature regime with algebraically decaying correlations is naturally explained as the regime in which all $O(2)$-invariant fields are irrelevant and the theory is conformal at large distances. Since $\Delta_\sigma=1/16$ at $b^2=1$, spin-spin correlations decay as $|x|^{-1/4}$ at $T_{BKT}$. The $XY$ model provides the basic illustration of the general fact that continuous symmetries do not break spontaneously in two dimensions \cite{MW,Hohenberg,Coleman_goldstone}.

It is known from exact lattice results \cite{Kaufman} that the two-dimensional Ising model corresponds to the theory of a free real fermion. Hence, the action (\ref{Thirring}) with $b^2=1/2$ describes the scaling limit of two decoupled Ising models, each with central charge $c=1/2$, energy density $\varepsilon_i=\psi_i\bar{\psi}_i$ and order field $\sigma_i$ ($i=1,2$). The conformal dimensions $\Delta_\varepsilon=1/2$ and $\Delta_{\sigma_\pm}=1/8$ at this decoupling point coincide with the dimensions $\Delta_{\varepsilon_i}$ and $\Delta_{\sigma_1\sigma_2}=2\Delta_{\sigma_i}$, respectively, which follow from exact results for the lattice Ising model \cite{Onsager,Yang}.

For later use we are also interested in the chiral (i.e. depending on $z$ or $\bar{z}$ only) fields which satisfy (\ref{quantization}) and have lowest charge $m=\pm 1$; they are $\eta_\pm=V_{\pm 1/2b}$ and $\bar{\eta}_\pm=\bar{V}_{\pm 1/2b}$, with $\Delta_{\eta_\pm}=\bar{\Delta}_{\bar{\eta}_\pm}=1/4b^2$. For $b$ generic the spin of these fields is neither integer nor half-integer, and for this reason we call them ``parafermionic'' fields \cite{FK_paraf,FZ_paraf}.

\subsection{$c<1$}
\label{c2}
In the case of central charge smaller than 1, which corresponds to $Q$ real, it is useful to parameterize $Q$ as
\EQ
Q=\beta^{-1}-\beta\,,\hspace{1cm}\beta^2=t/(t+1)\,,
\label{beta-t}
\EN
and to introduce the notation
\EQ
\Phi_{\mu,\nu}(z)=V_{p_{\mu,\nu}}(z)\,,\hspace{1cm}\bar{\Phi}_{\mu,\nu}(\bar{z})=\bar{V}_{p_{\mu,\nu}}(\bar{z})\,,
\label{phi}
\EN
where $p_{\mu,\nu}=\frac12[(1-\mu)\beta^{-1}-(1-\nu)\beta]$. Then the central charge (\ref{c}) can be rewritten as 
\EQ
c=1-\frac{6}{t(t+1)}\,,
\label{c_t}
\EN
and (\ref{deltap}) gives for the single non-vanishing conformal dimension of the fields (\ref{phi}) the result
\EQ
\Delta_{\mu,\nu}=\frac{[(t+1)\mu-t\nu]^2-1}{4t(t+1)}\,.
\label{deltamunu}
\EN
It is known from \cite{BPZ} that when the indices $\mu$ and $\nu$ take positive integer values $m$ and $n$ the dimensions (\ref{deltamunu}) are those of the degenerate primary fields associated to the algebra (\ref{Virasoro}) for $c<1$, and that the differential equations satisfied by multi-point correlators with degenerate fields imply the degenerate--non-degenerate and degenerate-degenerate OPE's
\bea
\Phi_{m,n}\cdot\Phi_{\mu,\nu}&=&\sum_{k=0}^{m-1}\,\sum_{l=0}^{n-1}\left[\Phi_{\mu-m+1+2k,\nu-n+1+2l}\right]\,,
\label{dnd}\\
\Phi_{m_1,n_1}\cdot\Phi_{m_2,n_2}&=&\sum_{k=0}^{\textrm{min}(m_1,m_2)-1}\,\sum_{l=0}^{\textrm{min}(n_1,n_2)-1}\left[\Phi_{|m_1-m_2|+1+2k,|n_1-n_2|+1+2l}\right]\,,
\label{dd}
\eea
with similar relations for the fields $\bar{\Phi}_{\mu,\nu}$.

The energy density field $\varepsilon$ is consistently defined if the OPE $\varepsilon\cdot\varepsilon$ does not produce fields (other than the identity) with scaling dimension smaller than $X_\varepsilon=2\Delta_\varepsilon$ and, within the present knowledge of theories with $c<1$, this seems to require that $\varepsilon$ is a degenerate field. We further restrict our analysis to theories which reduce to the Ising model for $c=1/2$, i.e. have $\Delta_\varepsilon=1/2$ for $t=3$. Then (\ref{deltamunu}) and (\ref{dd}) leave us with $\Phi_{2,1}\bar{\Phi}_{2,1}$ and $\Phi_{1,3}\bar{\Phi}_{1,3}$ as possible choices for $\varepsilon$. For the two choices, the OPE's (\ref{dnd}) and (\ref{dd}) can be used to select the fields mutually local with $\varepsilon$ and, among these, the parafermion $\eta(z)$ and the order field $\sigma(x)$, which has to satisfy
\EQ
\varepsilon\cdot\sigma=\sigma+\cdots\,
\label{epsilonsigma}
\EN
on symmetry grounds. 

For $\varepsilon=\Phi_{2,1}\bar{\Phi}_{2,1}$ this analysis \cite{paraf} yields $\eta=\Phi_{1,3}$ and $\sigma=\Phi_{1/2,0}\bar{\Phi}_{1/2,0}$. In this case the OPE $\eta\cdot\sigma$ produces a field $\mu$ with the same conformal dimensions of $\sigma$; the reason we do not identify $\mu$ with $\sigma$ is that we expect $\eta$ to be charged with respect to the symmetry $G$ (to be identified), as we saw for $c=1$, so that $\sigma$ and $\mu$ cannot transform in the same way. The field $\mu$ is called ``disorder'' field and its presence is the signature of an order-disorder ``duality''. For $\varepsilon=\Phi_{1,3}\bar{\Phi}_{1,3}$ one finds $\eta=\Phi_{2,1}$, while (\ref{epsilonsigma}) does not help for the identification of $\sigma$.

The two choices of $\varepsilon$ define critical lines parameterized by $t$ in the space of $c<1$ conformal field theory. Before discussing them in the next section, we mention that $c<1$ conformal field theory also allows for discrete series of critical points with special properties. Indeed it can be seen that for rational values of the parameter $\beta^2$ in (\ref{beta-t}) the OPE (\ref{dd}) closes on a {\it finite} number of degenerate primaries; one can then build consistent conformal theories containing a finite number of primaries and their descendants which are known as ``minimal models'' \cite{BPZ}. In particular, for $t=3,4,\ldots$, the conformal dimensions $\Delta_{m,n}$ of the primary fields are given by (\ref{deltamunu}) with $m=1,\ldots,t-1$, $n=1,\ldots,t$, and are all non-negative \cite{FQS}; this set of minimal models describes the critical ($t=3$) and multicritical ($t>3$) points associated to $Z_2$ symmetry\footnote{The scaling dimensions of primaries in the minimal model with $t=3$ (Ising) are $\Delta_{1,1}=\Delta_{2,3}=0$, $\Delta_{1,2}=\Delta_{2,2}=1/16$ and $\Delta_{1,3}=\Delta_{2,1}=1/2$, corresponding to the identity, $\sigma$ and $\varepsilon$, respectively.} \cite{Zamo_multicritical,Huse}. The model with $t=5$, restricted on a smaller set of primaries \cite{Cardy_modular,CIZ}, corresponds to the critical point of the three-state Potts model \cite{Dotsenko}, as we will also see in the next section\footnote{We also mention that the precise status of degenerate fields is the object of current studies in the so-called ``logarithmic'' cases \cite{Gurarie}, where the coincidence of the dimensions of two fields may lead to logarithmic corrections to the algebraic behavior (\ref{power}) of correlators (see \cite{GJRSV} for a recent review and a list of references). Normally, the collision of scaling dimensions and the associated logarithmic corrections can be observed at special points on critical lines with ordinary power law correlations (see e.g. \cite{DMM}).}.

\begin{figure}
\begin{center}
\includegraphics[width=7cm]{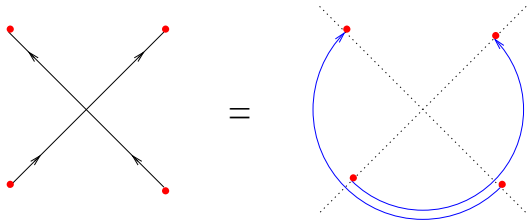}
\caption{Pictorial illustration of equation (\ref{phase}) in $(1+1)$-dimensional space-time.}
\label{paraf_phases}
\end{center} 
\end{figure}

\section{Scale invariant scattering and critical exponents}
\label{scale}
\subsection{Scattering and statistics}
We have seen in section~\ref{general} that the canonical critical exponents are determined by the scaling dimensions $X_\varepsilon=2\Delta_\varepsilon$ and $X_\sigma=2\Delta_\sigma$ of the energy density and order fields. We also introduced the $q$-state Potts and $n$-vector models and saw that their continuations to real values of $q$ and $n$ are related to percolation and non-intersecting walks, respectively. Since the Ising model, which corresponds to $q=2$ and $n=1$, has central charge $c=1/2$, these continuations should produce lines of critical points contained, entirely or in part, within the space of theories with $c\leq 1$ discussed in the previous section. In order to identify $\Delta_\varepsilon$ and $\Delta_\sigma$ in the two cases we need to relate these conformal field theories with the symmetries $S_q$ and $O(n)$ which characterize the two universality classes. Following \cite{paraf}, we will do this using the particle description introduced in section~\ref{particles}. The results we will obtain for the exponents coincide with those originally determined within the lattice Coulomb gas framework \cite{Nienhuis} and cast in the framework of conformal field theory in \cite{DF}. 

The elementary excitations of a scale invariant relativistic (1+1)-dimensional theory are massless particles with energy and momentum related as $p=e>0$ (right movers) or $p=-e<0$ (left movers). These particles are created by the chiral\footnote{Recalling (\ref{shift}) we have $\langle 0|\Phi(x)|p\rangle$ proportional to $e^{ipz}$ for a right mover and to $e^{ip\bar{z}}$ for a left mover.} fields $\eta$, with $\bar{\Delta}_\eta=0$, if right movers, and $\bar{\eta}$, with $\Delta_{\bar{\eta}}=0$, if left movers; one also has $s_\eta=-s_{\bar{\eta}}=\Delta_{{\eta}}=\bar{\Delta}_{\bar{\eta}}$ for the spin of the fields. Due to the absence of dimensional parameters, the dimensionless elastic scattering amplitude, that we denote by ${\cal S}$, of a right-mover with a left-mover cannot depend on the variable (\ref{e2}), which is the only relativistic invariant in the process and is dimensionful. The fact that ${\cal S}$ does not depend on the particle energy and, by unitarity, is a constant phase means that the two particles have no dynamical interaction. The scattering process, however, involves exchanging the position of the two particles on the line, so that scattering entails in general a statistical factor; ${\cal S}$ will be 1 for bosons ($\Delta_\eta$ integer), $-1$ for fermions ($\Delta_\eta$ half-integer), and a more general phase when the fields $\eta$ and $\bar{\eta}$ are parafermionic. 
Since in absence of dynamical interaction the passage from the initial to the final state can also be realized by $\pi$-rotations (see Fig.~\ref{paraf_phases}) ruled by the Euclidean spin, the right-left scattering (statistical) phase can be written as \cite{paraf}
\EQ
{\cal S}=e^{-i\pi(s_\eta-s_{\bar{\eta}})}=e^{-2i\pi\Delta_\eta}\,.
\label{phase}
\EN

In general the particles and the associated chiral fields form multiplets transforming under a representation of the symmetry of the theory. We denote by $\eta_a$ and $\bar{\eta}_a$ the components of such multiplets, and encode the general right-left scattering in the symbolic relation
\EQ
\eta_a\circ\bar{\eta}_b=\sum_{c,d}{\cal S}_{ab}^{cd}\,\,\,\bar\eta_d\circ\eta_c\,,
\label{algebra}
\EN
where ${\cal S}_{ab}^{cd}$ are scattering amplitudes (Fig.\,\ref{fpu_elastic}); then (\ref{phase}) is obtained upon diagonalization of (\ref{algebra}). Unitarity yields
\EQ
\sum_{e,f}{\cal S}_{ab}^{ef}\left[{\cal S}_{ef}^{cd}\right]^*=\delta_a^c\delta_b^d\,,
\label{massless_unitarity}
\EN
while crossing symmetry (\ref{crossing0}) can be written as
\EQ
{\cal S}_{ab}^{cd}=\left[{\cal S}_{\bar{d}a}^{\bar{b}c}\right]^*
\label{massless_crossing}
\EN
recalling real analyticity. We can now proceed to the analysis of specific cases.

\begin{figure}
\begin{center}
\includegraphics[width=9cm]{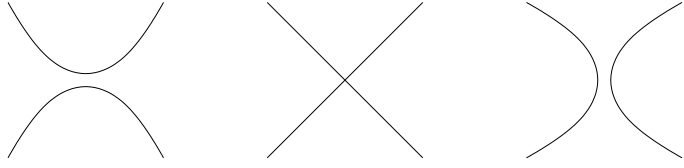}
\caption{Amplitudes ${\cal S}_1$, ${\cal S}_2$ and ${\cal S}_3$ of the $O(n)$-invariant theory.}
\label{On_ampl}
\end{center} 
\end{figure}

\subsection{$O(n)$ symmetry and non-intersecting walks}
\label{o(n)}
In this case the symmetry is naturally represented by vector multiplets $\eta_a$, $\bar{\eta}_a$, $a=1,\ldots,n$, (\ref{algebra}) takes the $O(n)$-covariant form (Fig.~\ref{On_ampl})
\EQ
\eta_a\circ\bar{\eta}_b=\delta_{ab}\,{\cal S}_1\sum_{c=1}^n\bar\eta_c\circ\eta_c+
{\cal S}_2\,\,\bar\eta_b\circ\eta_a+{\cal S}_3\,\,\bar\eta_a\circ\eta_b\,,
\label{On_algebra}
\EN
(\ref{massless_crossing}) gives ${\cal S}_1={\cal S}_3^*\equiv\rho_1e^{i\varphi}$, ${\cal S}_2={\cal S}_2^*\equiv\rho_2$, with $\rho_1$ non-negative and $\rho_2$ real\footnote{The phase $\varphi$ should not be confused with the field in (\ref{free}).}, and (\ref{massless_unitarity}) is then rewritten as
\bea
&& \rho_1^2+\rho_2^2=1\,,
\label{uni1}\\
&& \rho_1\rho_2\cos\varphi=0\,,
\label{uni2}\\
&& n\rho_1^2+2\rho_1\rho_2\cos\varphi+2\rho_1^2\cos 2\varphi=0\,.
\label{uni3}
\eea
There are three ways of satisfying (\ref{uni2}). The first possibility is that $\cos\varphi=0$, so that (\ref{uni3}) fixes $n=2$, a case to which we will come back in a moment. The second possibility is that $\rho_1=0$, but then we are left with ${\cal S}_2=\pm 1$ as the only non-vanishing amplitude, i.e. with $n$ decoupled free bosons or fermions. Hence, the only non-trivial case with a continuous $n$-dependence is
\EQ
\rho_2=0\,,\hspace{1cm}\rho_1=1\,,\hspace{1cm}n=-2\cos 2\varphi\in(-2,2)\,.
\label{On}
\EN
Thinking of particle trajectories as walks of $n$ different colors on the Euclidean plane, the result ${\cal S}_2=0$ amounts to a no-crossing condition for the walks (Fig.~\ref{On_ampl}), as first observed in \cite{selfavoiding} in the study of the off-critical case.

It follows from (\ref{On_algebra}) and (\ref{On}) that 
\EQ
\sum_{a=1}^n\eta_a\circ\bar{\eta}_a={\cal S}\,\,\sum_{a=1}^n\bar\eta_a\circ\eta_a\,,
\label{algebra_On}
\EN
\EQ
{\cal S}=n{\cal S}_1+{\cal S}_2+{\cal S}_3=-e^{3i\varphi}\,,
\label{S_On}
\EN
and (\ref{phase}) then relates $\varphi$ to $\Delta_\eta$, up to the $2\pi k$ ambiguity involved in the comparison of phases. The relations (\ref{On}) and (\ref{S_On}) consistently give $n=1$ and the fermionic (Ising) value ${\cal S}=-1$ when $\varphi=2\pi k/3$, $k=1,2\,$(mod\,3). Looking for matching with one of the two $c<1$ critical lines identified in the previous section, we take that with $\varepsilon=\Phi_{1,3}\bar{\Phi}_{1,3}$ and $\eta=\Phi_{2,1}$, since the order-disorder duality exhibited by the other is typical of the Potts model \cite{Potts,Wu}. Then we use $\Delta_\varepsilon=\Delta_{1,3}=1$ at $c=1$ and $\Delta_\eta=1/4$ for $\Delta_\varepsilon=1$ (i.e. for $b^2=1$) on the $O(2)$ line of the previous section to fix $\varphi=-\frac{2\pi}{3}(\Delta_\eta+1/2)$\,(mod\,$2\pi$); substituting $\Delta_\eta=\Delta_{2,1}$ and plugging into (\ref{On}) we finally obtain
\EQ
n=2\cos\frac{\pi}{t}\,,
\label{n}
\EN
where $t$ parameterizes the central charge through (\ref{c_t}). We see that the $n=2$ endpoint of the solution (\ref{On}) corresponds to the BKT transition point $b^2=1$ on the $O(2)$ line with $c=1$ of the previous section. Hence we conclude that (\ref{On}) gives $\Delta_\sigma=1/16$ not only at $c=1/2$ but also at $c=1$; using these conditions to determine the indices (assumed $t$-independent) in (\ref{deltamunu}) gives $\Delta_\sigma=\Delta_{1/2,0}$ as the only positive solution within the range $c\in(0,1)$. 

Coming back to the solution of (\ref{uni1}-\ref{uni3}) with $n=2$, it can be parameterized as
\EQ
\rho_1=\sin\alpha\,,\hspace{1cm}\rho_2=\cos\alpha\,,\hspace{1cm}\varphi=-\pi/2\,,\label{O2}
\EN
where we now allow for $\rho_1$ negative. The phase in (\ref{algebra_On}) becomes
\EQ
{\cal S}=2{\cal S}_1+{\cal S}_2+{\cal S}_3=e^{-i\alpha}\,,
\label{S_O2}
\EN
and has to be equated to (\ref{phase}) with $\Delta_\eta$ which, as seen in the previous section, takes the value $1/4b^2$ along the $O(2)$-invariant critical line parameterized by $b^2$; this gives 
\EQ
\alpha=\frac{\pi}{2b^2}\,,
\label{alpha}
\EN
a relation that we will exploit later on.
\begin{figure}
\begin{center}
\includegraphics[width=10cm]{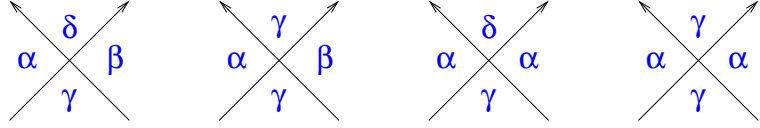}
\caption{Amplitudes ${\cal S}_0$, ${\cal S}_1$, ${\cal S}_2$ and ${\cal S}_3$ of the $S_q$-invariant theory.}
\label{potts_ampl}
\end{center} 
\end{figure}

\subsection{Permutational symmetry and cluster boundaries}
\label{sq}
In view of the cluster representation ({\ref{FK}) of the Potts partition function, we associate in this case the particle trajectories to the boundaries separating different clusters, denote by $\eta_{\alpha\beta}$, $\bar{\eta}_{\alpha\beta}$ ($\alpha,\beta=1,\ldots,q$; $\alpha\neq\beta$) the chiral fields corresponding to the trajectories separating a cluster of color $\alpha$ from one of color $\beta$, and use the permutational symmetry of the colors to write (\ref{algebra}) in the form\footnote{This color structure for the scattering problem was first observed in \cite{CZ} within the study of the model below $T_c$, with the particles associated to kinks interpolating between degenerate ferromagnetic vacua.} (Fig.~\ref{potts_ampl})
\bea
\eta_{\alpha\gamma}\circ\bar{\eta}_{\gamma\beta}=(1-\delta_{\alpha\beta})
\left[{\cal S}_0\sum_{\delta\neq\gamma}\bar\eta_{\alpha\delta}\circ\eta_{\delta\beta}+
{\cal S}_1\,\,\bar\eta_{\alpha\gamma}\circ\eta_{\gamma\beta}\right]&&\nonumber\\
+ \delta_{\alpha\beta}\left[
{\cal S}_2\sum_{\delta\neq\gamma}\bar\eta_{\alpha\delta}\circ\eta_{\delta\alpha}+  
{\cal S}_3\,\,\bar\eta_{\alpha\gamma}\circ\eta_{\gamma\alpha}\right]&&.
\label{algebra_Sq}
\eea
Crossing symmetry enables us to write
\EQ
{\cal S}_0={\cal S}_0^*\equiv\rho_0\,,\hspace{1cm}{\cal S}_1={\cal S}_2^*\equiv\rho e^{i\varphi}\,,\hspace{1cm}{\cal S}_3={\cal S}_3^*\equiv\rho_3\,,
\EN
with $\rho_0$ and $\rho_3$ real and $\rho$ non-negative, so that the unitarity equations take the form
\bea
&&(q-3)\rho_0^2+\rho^2=1\,,\nonumber\\
&&(q-4)\rho_0^2+2\rho_0\rho\cos\varphi=0\,,\nonumber\\
&&(q-2)\rho^2+\rho_3^2=1\,,\nonumber\\
&&(q-3)\rho^2+2\rho\rho_3\cos\varphi=0\,.\nonumber
\eea
The solution 
\EQ
\rho_0=-1\,,\hspace{1cm}\rho=\sqrt{4-q}\,,\hspace{1cm}2\cos\varphi=-\sqrt{4-q}\,,\hspace{1cm}\rho_3=q-3\,,
\label{potts_solution}
\EN
is the one\footnote{We use the condition $\rho_3=-1$ at $q=2$ (Ising) to fix the sign of $\rho_3$.} which identifies a $S_q$-invariant critical line with $q\in(0,4)$, in agreement with the fact, known exactly from lattice studies \cite{Baxter}, that the Potts phase transition becomes of the first order for $q>4$, thus preventing a continuum limit. At $q=2$ the only physical amplitude is ${\cal S}_3=-1$, as needed for the Ising model, and the solution (\ref{potts_solution}) must be associated to the remaining theory with $c\leq 1$ of the previous section, that with $\varepsilon=\Phi_{2,1}\bar{\Phi}_{2,1}$ and $\eta=\Phi_{1,3}$; $c=1$ should correspond to the maximal value of $q$, which we saw is 4. It follows from (\ref{algebra_Sq}) and (\ref{potts_solution}) that
\EQ
\sum_\gamma\eta_{\alpha\gamma}\circ\bar{\eta}_{\gamma\alpha}=
{\cal S}\,\,\sum_\gamma\bar\eta_{\alpha\gamma}\circ\eta_{\gamma\alpha}\,,
\EN
\EQ
{\cal S}={\cal S}_3+(q-2){\cal S}_2=e^{-4i\varphi}\,.
\label{S_Sq}
\EN
Using (\ref{phase}) and the requirement that $\Delta_\eta=\Delta_{1,3}$ equals $1/2$ and 1 for $q=2$ and 4, respectively, one finds $\varphi=\pi(1+\frac12\Delta_{1,3})$ (mod\,$2\pi$) and, upon substitution in (\ref{potts_solution}), 
\EQ
\sqrt{q}=2\sin\frac{\pi(t-1)}{2(t+1)}\,.
\label{q}
\EN
Table~1 summarizes the results for the three critical lines we have discussed in this section. 

\begin{table}
\begin{center}
\begin{tabular}{|l||c|c|c|c|}
\hline
Symmetry & $c$ & $\Delta_{\varepsilon}$ & $\Delta_\eta$ & $\Delta_{\sigma}$ \\
\hline
$O(2)$ & $1$ & $b^2$ & $\frac{1}{4b^2}$ & $\frac{1}{16b^2}$ \\
$O(n),\hspace{.4cm}n=2\cos\frac{\pi}{t}$ & $1-\frac{6}{t(t+1)}$ & $\Delta_{1,3}$ & $\Delta_{2,1}$ & $\Delta_{1/2,0}$ \\
$S_q,\hspace{.4cm}\sqrt{q}=2\sin\frac{\pi(t-1)}{2(t+1)}$ & $1-\frac{6}{t(t+1)}$ & $\Delta_{2,1}$ & $\Delta_{1,3}$ & $\Delta_{1/2,0}$ \\
\hline
\end{tabular}
\caption{Central charge and conformal dimensions along the three critical lines discussed in the text; $\Delta_{\mu,\nu}$ is given by (\ref{deltamunu}). The dimensions $\Delta_\varepsilon$ and $\Delta_\sigma$ of the energy density and order fields determine the canonical critical exponents.} 
\label{table1}
\end{center}
\end{table}

\section{Away from criticality}
\label{away}

\subsection{Integrability}
\label{integrability}
The fact that two-dimensional scale invariant theories yield exact results (in particular the critical exponents) has to be traced back to the infinite-dimensional conformal symmetry codified by the algebra (\ref{Virasoro}), which in turn implies the existence of infinitely many conservation laws. A conservation equation in two dimensions takes the form 
\EQ
\bar{\partial}T_{s+1}=\partial\Theta_{s-1}\,,
\label{conservation}
\EN
where the subscripts specify the Euclidean spin of the fields. At criticality, the descendants of the identity with conformal dimensions $(\Delta,\bar{\Delta})=(s+1,0)$ yield infinitely many local fields $T_{s+1}$ satisfying (\ref{conservation}) with zero on the r.h.s; hence, that of energy and momentum ($s=1$, Eq.~(\ref{Tconservation})) is only one among infinitely many conservation equations. When we move away from criticality and consider a theory with action (\ref{scaling}), $\bar{\partial}T_{s+1}$ no longer vanishes, but for $s\neq 1$ has no need to become a total derivative with respect to $z$. In other words, non-trivial (i.e. other than energy and momentum) conserved quantities are normally lost away from criticality. Instead of (\ref{conservation}) we will have in general the expansion \cite{Taniguchi}
$\bar{\partial}T_{s+1}=\sum_{n=1}^N\tau^nA^{(n)}_s$, 
where $\tau$ with positive dimensions $(1-\Delta_\varepsilon,1-\Delta_\varepsilon)$ is the coupling in (\ref{scaling}), and $A^{(n)}_s$ are fields with dimensions $(s+1-n(1-\Delta_\varepsilon),1-n(1-\Delta_\varepsilon))$. We consider theories with a spectrum of dimensions discrete\footnote{While the assumption of discrete spectrum of dimensions covers the main cases and is certainly sufficient for our purposes, it is interesting to mention that lattice models pointing to a continuous spectrum of dimensions have been investigated \cite{VJS}.} and bounded from below. Then $N$ not only has to be finite, to avoid fields $A^{(n)}_s$ with arbitrarily large negative dimensions, but generically has to be 1, since only $A^{(1)}_s\equiv A_s$ can be identified with a field certainly present in the theory, i.e. the descendant of $\varepsilon$ with dimensions $(\Delta_\varepsilon+s,\Delta_\varepsilon)$, without special conditions on $\Delta_\varepsilon$. Hence, in the generic case we are left with
\EQ
\bar{\partial}T_{s+1}=\tau\,A_s\,,
\label{conservation1}
\EN
which is a conservation equation if $A_s$ is $\partial$ of another field; of course we exclude the case in which $T_{s+1}$ is itself $\partial$ of another field. Let us call $D_{T_{s+1}}$ the number of linearly independent non-derivative fields $T_{s+1}$, i.e. the dimension of the space of level $s+1$ descendants of the identity which are not $\partial$ of other fields, and $D_{A_s}$ the dimension of the space of level $s$ non-derivative descendants of $\varepsilon$; then (\ref{conservation1}) is a conservation law if
\EQ
D_{A_s}<D_{T_{s+1}}\,.
\label{counting}
\EN
Such a {\it counting argument} \cite{Taniguchi} yields a sufficient condition for the existence of off-critical conserved quantities through the comparison of dimensionalities of subspaces in the space of fields; the structure of these subspaces has been described in section~\ref{basic}. For $s=1$ we have ${D}_{A_1}=0$ (the only level 1 field is $L_{-1}\varepsilon=\partial\varepsilon$) and $D_{T_{2}}=1$ ($L_{-2}I=T$), so that (\ref{counting}) is always satisfied and conservation of energy and momentum recovered. For $s>1$, as expected, (\ref{counting}) is violated in general. For example, for $s=3$ one has $D_{A_3}=D_{T_{4}}=1$, since $L_{-3}\varepsilon$ and $L_{-2}^2I$ are the only fields allowed\footnote{The derivative character of $L_{-4}I$ follows from the fact that (\ref{TPhi}) with $\Phi=I$ reduces to a Taylor expansion for $T(z)$.}. However, if $\varepsilon$ is a primary degenerate (in the sense defined in section~\ref{basic}) at level 3, $D_{A_3}$ reduces to 0 and (\ref{counting}) is satisfied. Hence we see that, remarkably, if the field which breaks conformal invariance is degenerate, the reduced dimensionality of the descendant subspaces leaves room for (\ref{counting}) to be fulfilled with $s>1$, i.e. for the existence of non-trivial conserved quantities. The systematic analysis performed in \cite{Taniguchi} shows that the relevant degenerate fields $\phi_{1,3}$, $\phi_{1,2}$ and $\phi_{2,1}$ allow for non-trivial conserved quantities when one of them is added to a conformal field theory with $c<1$; here we are introducing the notation
\EQ
\phi_{\mu,\nu}(x)=\Phi_{\mu,\nu}(z)\bar{\Phi}_{\mu,\nu}(\bar{z})
\label{scalar_primaries}
\EN
for scalar primaries obtained from the fields (\ref{phi}).

\begin{figure}
\begin{center}
\includegraphics[width=12cm]{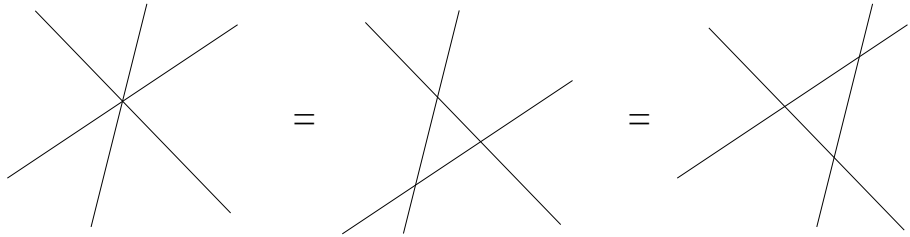}
\caption{Factorization of scattering amplitudes in integrable field theories.}
\label{fpu_factorization}
\end{center} 
\end{figure}

While the sufficient condition (\ref{counting}) turns out to yield a finite number of conservation laws, it is expected that these are a subset of an infinite series which leads to the exact solvability (or {\it integrability}) of the off-critical theory. Ultimately, this expectation is confirmed by the existence of exact $S$-matrix solutions, as we are going to see. Anticipating this step and recalling Table~\ref{table1}, we observe that the above results of the counting argument mean that the two-dimensional $q$-state Potts ($q\in(0,4)$) and $n$-vector ($n\in(-2,2)$) models are exactly solvable in the scaling limit away from criticality. The same conclusion follows for the Ising model at $T=T_c$ in presence of a magnetic field (the magnetic field amounts to the addition of $H\sum_i\sigma_i$ to the lattice Hamiltonian (\ref{ising})). The scaling limit at $T_c$ is obtained adding to the conformal theory with central charge $c=1/2$ the field $\sigma=\phi_{1/2,0}=\phi_{1,2}$; the last equality follows from (\ref{deltamunu}), which for $t=3$ gives $\Delta_{1/2,0}=\Delta_{1,2}=1/16$. This is one of many examples of models which are not solved on the lattice but become solvable in the scaling limit.

If a field theory possesses infinitely many conserved quantities its $S$-matrix drastically simplifies. First of all the infinitely many conservation equations force the final state of a scattering process to be kinematically identical (same number of particles, same masses, same momenta) to the initial one ({\it complete elasticity}); only reshuffling of charges is allowed. The second simplification can be understood through the following argument \cite{SW}. A single-particle state $|p\rangle$ is eigenstate of a spin $s$ conserved quantity with eigenvalue proportional to $(p+e)^s$. If the particle is located around $x_1=0$ at $t=0$ it can be described by the wave packet $\int dk f(k)e^{ikx_1}$, with $f(k)$ peaked around $p$. Stationary phase approximation then shows that the action on the wave packet of the translation operator $e^{-iPa}$ moves it to $x_1=a$, no matter the value of $p$. If, on the other hand, we repeat the operation replacing the momentum operator $P$ with the spatial component of a spin $s$ conserved quantity, this produces a factor $e^{iak^s}$ and the stationary phase approximation now yields a displacement proportional to $p^{s-1}$. This means that non-trivial conserved quantities ($s>1$) act on particles as generators of momentum-dependent spatial displacements. This fact can be used in a multiparticle scattering to make arbitrarily distant in space-time the two-body subprocesses, as well as to change their chronological order (Fig.~\ref{fpu_factorization}). Since the conserved quantities commute with time evolution, we conclude that $n$-particle scattering amplitudes {\it factorize} into the product of $n(n-1)/2$ two-particle amplitudes, and that 
\EQ
{\cal S}_{12}{\cal S}_{13}{\cal S}_{23}={\cal S}_{23}{\cal S}_{13}{\cal S}_{12}\,,
\label{factorization}
\EN
where ${\cal S}_{ij}$ is a short notation for the amplitude of the particles with momenta $p_i$ and $p_j$. Let us stress that Fig.~\ref{fpu_factorization}, and then factorization, only makes sense in $1+1$ dimensions, since two particles with different momenta necessarily meet on a line. As we will see below, when added to unitarity and crossing symmetry, complete elasticity and factorization allow the exact determination of the $S$-matrix.

\begin{figure}
\begin{center}
\includegraphics[width=14cm]{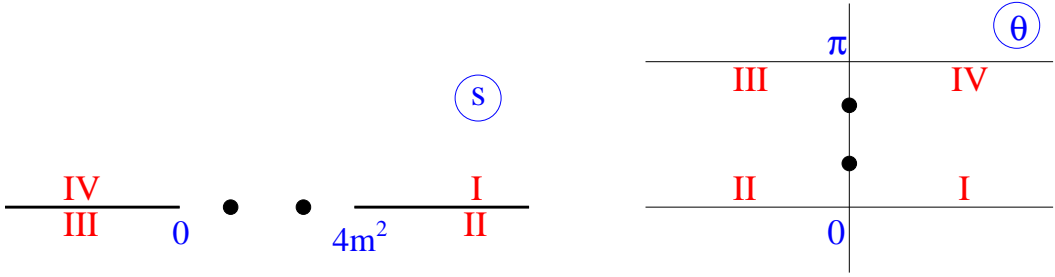}
\caption{The cut $s$-plane is mapped onto the cut-free $\theta$-plane through the relation (\ref{s_theta}).}
\label{fpu_s_theta}
\end{center} 
\end{figure}

Factorization of scattering amplitudes in integrable field theories reduces the determination of the $S$-matrix to that of the two-particle amplitudes of Fig.~\ref{fpu_elastic} \cite{ZZ}. Complete elasticity of the scattering implies that these amplitudes exhibit in the $s$-plane only the elastic unitarity cut and its crossing-symmetric image (Fig.~\ref{fpu_s_theta}). It is convenient to introduce the rapidity parameterization 
\EQ
(e,p)=(m\cosh\theta,m\sinh\theta)
\label{rapidity}
\EN
for the momentum and energy of a particle of mass $m$; a Lorentz transformation corresponds to a shift of the rapidity, so that relativistic invariant quantities depend on rapidity differences. In particular, the square of the center of mass energy (\ref{e2}) becomes 
\EQ
s=4m^2\cosh^2\frac{\theta_{12}}{2}
\label{s_theta}
\EN
 for two particles of equal mass and rapidities $\theta_1$, $\theta_2$ ($\theta_{12}\equiv\theta_1-\theta_2$), and we will use the notation ${\cal S}_{ab}^{cd}(s)=S_{ab}^{cd}(\theta)$ for the amplitudes. The fact that different values of $\theta$ correspond to the same value of $s$ can be exploited to map different sheets of the $s$-plane onto different regions of the $\theta$-plane, the physical sheet corresponding to the physical strip Im$\,\theta\in(0,\pi)$ (Fig.~\ref{fpu_s_theta}); the amplitudes $S_{ab}^{cd}(\theta)$ are then ``meromorphic'' functions, i.e. allow for poles as the only singularities in the complex $\theta$-plane. Due to complete elasticity, the unitarity condition (\ref{unitarity0}) holds at any energy, and in the rapidity variables becomes
\EQ
\sum_{e,f}S_{ab}^{ef}(\theta)S_{ef}^{cd}(-\theta)=\delta_{ac}\delta_{bd}\,;
\label{unitarity}
\EN
the crossing relation (\ref{crossing0}) translates into
\EQ
S_{ab}^{cd}(\theta)=S_{\bar{d}a}^{\bar{b}c}(i\pi-\theta)\,.
\label{crossing}
\EN

\begin{figure}
\begin{center}
\includegraphics[width=1.5cm]{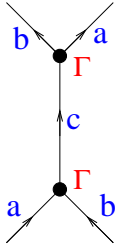}
\caption{A bound state pole of the scattering amplitude.}
\label{fpu_pole}
\end{center} 
\end{figure}

\subsection{Purely trasmissive scattering}
\label{pure}
When looking for solutions of the factorized scattering problem the simplest possibility is to consider purely transmissive scattering, i.e. $c=a$ and $d=b$ in Fig.~\ref{fpu_elastic}. For this case we can set $S_{ab}^{ab}(\theta)\equiv S_{ab}(\theta)$ and (\ref{unitarity}), (\ref{crossing}) reduce to 
\EQ
S_{ab}(\theta)S_{ab}(-\theta)=1\,,
\label{diagonal1}
\EN
\EQ
S_{ab}(\theta)=S_{\bar{b}a}(i\pi-\theta)\,.
\label{diagonal2}
\EN
The solutions to (\ref{diagonal1}) in the space of functions meromorphic in $\theta$ and real analytic and polynomially bound in $s$ can be written in the form $\prod_\alpha f_\alpha(\theta)$, with
\EQ
f_\alpha(\theta)=\frac{\sinh[(\theta+i\pi\alpha)/2]}{\sinh[(\theta-i\pi\alpha)/2]}\,,\hspace{1cm}\alpha\in(0,2)\,.
\label{fa}
\EN
A pole 
\EQ
S_{ab}(\theta)\simeq i\frac{(\Gamma_{ab}^c)^2}{\theta-iu_{ab}^c}\,,\hspace{1cm}\theta\simeq iu_{ab}^c
\label{pole}
\EN
in the physical strip corresponds to a bound state with mass $m_c=2m\cos(u_{ab}^c/2)$ (Fig.~\ref{fpu_pole}). Up to the factor of $i$, such a direct channel pole has a positive residue; through (\ref{diagonal2}) it induces a negative residue pole in the crossed channel.

Trivial examples of purely transmissive scattering are provided by free theories. It follows from the results of section~\ref{conformal} that these include the relevant case of the scaling Ising model which, in absence of magnetic field, corresponds to the theory of a free massive real fermion, and then to $S(\theta)=-1$. This amplitude describes both scaling regimes  above and below $T_c$, since the difference in the action is only a change of sign of the mass term $\tau\psi\bar{\psi}$. Below $T_c$, however, there are two degenerate vacua and the free fermions corresponds to the kinks which interpolate between them, which in $(1+1)$ dimensions are stable for topological reasons and always provide the elementary excitations in presence of degenerate vacua. If $K_{\alpha\beta}$ denotes a kink going from the vacuum $\alpha$ to the vacuum $\beta$, in the Ising case we only have $K_{+-}$ and $K_{-+}$, and multi-kink states can only have the form $\cdots K_{+-}K_{-+}K_{+-}\cdots$, with the sequence fixed by the choice of the first vacuum. By symmetry there is a single two-kink amplitude, which can be depicted as the last one of Fig.~\ref{potts_ampl} and is equal to $-1$. The coincidence of the amplitudes above and below $T_c$ is the manifestation, within the particle description, of the duality of the Ising model well known from the lattice formulation \cite{ising_duality}.

To illustrate the interacting case we consider a theory invariant under a charge conjugation symmetry and whose elementary excitations are two conjugated particles that we denote by $A$ and $\bar{A}$; they have charge $q$ and $-q$, respectively. The simplest non-constant solution of (\ref{diagonal1}) and (\ref{diagonal2}) reads \cite{KS}
\EQ
S_{AA}(\theta)=f_{2/3}(\theta)\,,\hspace{1cm}S_{A\bar{A}}(\theta)=-f_{1/3}(\theta)\,.
\label{3potts}
\EN
The overall sign is chosen so that the pole at $\theta=2i\pi/3$ of $S_{AA}$ is positive, and then corresponds to a bound state with the same mass $m$ which does not introduce new particle species. Identifying this particle with $A$ would imply $q=0$; it then corresponds to $\bar{A}$, and we have $2q=-q$, i.e. identification of charge modulo $3q$. Hence, the amplitudes (\ref{3potts}) yield the exact solution of the simplest theory possessing $Z_3$ symmetry ($S_3$ taking into account also charge conjugation). The freedom to separately shift particle trajectories in integrable theories also yields the ``bootstrap'' equation (Fig.~\ref{fpu_bootstrap})
\EQ
S_{A\bar{A}}(\theta)=S_{A{A}}(\theta-i\pi/3)S_{A{A}}(\theta+i\pi/3)
\label{bootstrap}
\EN
in connection with the vertex $AA\to\bar{A}$; it is satisfied by (\ref{3potts}). 

\begin{figure}
\begin{center}
\includegraphics[width=5cm]{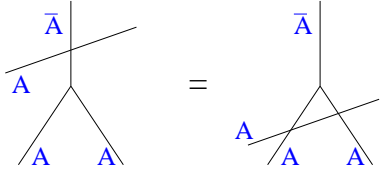}
\caption{A bootstrap equation.}
\label{fpu_bootstrap}
\end{center} 
\end{figure}

On universality grounds it is natural to identify (\ref{3potts}) to the scattering solution of the scaling three-state Potts model, whose integrability we argued above. The identification is correct \cite{Zamo_3potts,TBA} and in this as in the other cases the ultimate check is provided by a technique known as thermodynamic Bethe ansatz \cite{TBA}. In this approach one studies the thermodynamics of the gas of relativistic particles confined on a circle, and uses the scattering amplitudes to specify the interactions in the low-density limit. When the temperature of the gas becomes much larger than the mass of the particles the theory approaches the scale invariant limit and it is possible to determine the central charge starting from the scattering amplitudes.

The above solution in terms of $A$ and $\bar{A}$ corresponds to the scaling three-state Potts model above $T_c$. Below $T_c$ there are three degenerate vacua and the associated kinks, together with the symmetry, allow for the inequivalent amplitudes $S_1$, $S_2$ and $S_3$ corresponding to the last three in Fig.~\ref{potts_ampl}. The particle-kink correspondences $A\leftrightarrow K_{\alpha,\alpha+1}$, $\bar{A}\leftrightarrow K_{\alpha,\alpha-1}$, with indices taking integer values identified modulo 3, allow the identifications $S_{AA}=S_1$, $S_{A\bar{A}}=S_2$, and then the description of both regimes above and below $T_c$ in terms of the same amplitudes ($S_3=S_{A\bar{A}}^{\bar{A}A}=0$). This is expected, since the lattice $q$-state Potts model exhibits high-low temperature duality \cite{Potts,Wu}.

As another relevant example of purely transmissive theory we mention the Ising model at $T_c$ with a magnetic field, whose integrability we also discussed. The presence of the magnetic field leaves no internal symmetry, so that the mass spectrum has to be non-degenerate and the scattering transmissive. A detailed analysis of non-trivial conservation laws at bound state vertices was exploited in \cite{Taniguchi} to determine the full particle spectrum (which consists of eight particles with different masses) and the scattering amplitudes. The reader is referred to \cite{review} for a review of this solution and of its implications.

\subsection{In presence of backscattering}
\label{generic}
The case in which the scattering is factorized but not purely transmissive can be illustrated through the very relevant example of the sine-Gordon model. This corresponds to the action (\ref{scaling}) with the conformal invariant part ${\cal A}_{FP}$ given by the $c=1$ free massless bosonic action (\ref{free}), and $\varepsilon\propto\cos 2b\varphi$. It follows from the discussion of section~\ref{c1} that the model also admits the fermionic representation with (\ref{Thirring}) as the conformal part and $\varepsilon\propto(\psi_1\bar{\psi_1}+\psi_2\bar{\psi_2})$. We also discussed how the complex fermion (\ref{dirac}) and its conjugate $\Psi^*$ carry opposite units of the charge associated to the $U(1)$ symmetry; these fields create the elementary particle excitations of the theory, that we denote by $A$ and $\bar{A}$, respectively. In the bosonic language they correspond to the soliton and antisoliton which interpolate, in opposite directions, between adjacent minima of the periodic potential $\cos 2b\varphi$; the $U(1)$ charge then corresponds to a topologic charge.

The integrability of sine-Gordon model does not follow from the counting argument, which does not apply to $c=1$, but can be shown using more traditional methods of Lagrangian field theory (\cite{ZZ} and references therein). Hence the scattering solution requires the determination of the amplitudes $S_{ab}^{cd}(\theta)$ with indices taking values $A$ and $\bar{A}$. Charge conservation and charge conjugation symmetry allow for the three independent amplitudes
\EQ
S_{AA}^{AA}(\theta)\equiv S_0(\theta)\,,\hspace{1cm}
S_{A\bar{A}}^{A\bar{A}}(\theta)\equiv S_T(\theta)\,,\hspace{1cm}
S_{A\bar{A}}^{\bar{A}A}(\theta)\equiv S_R(\theta)\,,
\EN
which satisfy the crossing equations
\EQ
S_0(\theta)=S_T(i\pi-\theta)\,,\hspace{1cm}S_R(\theta)=S_R(i\pi-\theta)\,,
\label{sg_crossing}
\EN
and the unitarity equations
\bea
&& S_T(\theta)S_T(-\theta)+S_R(\theta)S_R(-\theta)=1\,,\label{sg_uni1}\\
&& S_T(\theta)S_R(-\theta)+S_R(\theta)S_T(-\theta)=0\,,\label{sg_uni2}\\
&& S_0(\theta)S_0(-\theta)=1\,.\label{sg_uni3}
\eea
The presence of the reflection amplitude $S_R$ makes non-trivial the factorization equations (\ref{factorization}) since, for fixed initial and final states, one has to sum over the different particle indices possible for the internal triangles in Fig.~\ref{fpu_factorization}. In particular, the process $A\bar{A}A\to AA\bar{A}$ gives
\EQ
S_T(\theta)S_0(\theta+\theta')S_R(\theta')+S_R(\theta)S_R(\theta+\theta')S_T(\theta')=S_R(\theta')S_T(\theta+\theta')S_0(\theta)\,.
\label{sg_fact}
\EN
For the theories in which the scattering is not purely transmissive the amplitudes cannot be written as products of basic blocks like (\ref{fa}), and their determination requires a non-trivial model dependent study of the functional equations. For sine-Gordon this can be found in \cite{ZZ} and leads to the minimal (i.e with the minimal number of poles) solution to (\ref{sg_crossing}-\ref{sg_fact})
\EQ
S_T(\theta)=-\frac{\sinh\frac{\pi\theta}{\xi}}
               {\sinh\left(\frac{\pi}{\xi}(\theta-i\pi)\right)}\,S_0(\theta)\,,\hspace{1cm}S_R(\theta)=-\frac{\sinh\frac{i\pi^2}{\xi}}
               {\sinh\left(\frac{\pi}{\xi}(\theta-i\pi)\right)}\,S_0(\theta)\,,
\label{sg_tr}
\EN
\EQ
S_0(\theta)=-\exp\left\{-i\int_0^\infty\frac{dx}{x}
\frac{\sinh\left[\frac{x}{2}\left(1-
\frac{\xi}{\pi}\right)\right]}{\sinh\frac{x\xi}{2\pi}\cosh\frac{x}{2}}
\sin\frac{\theta x}{\pi}\right\}\,;
\label{sg_s0}
\EN
the amplitude $S_0$ is initially found in the form of an infinite product of gamma functions, more suitable for dealing with the functional equations, and then exponentiated using the formula
\EQ
\Gamma(z)=\exp\left\{\int_0^\infty\frac{dt}{t}\left[\frac{e^{-tz}-e^{-t}}{1-e^{-t}}+(z-1)e^{-t}\right]\right\}\,.
\label{logamma}
\EN
The parameter $\xi$ allowed by the solution of the functional equations for the amplitudes must be related to the parameter $b$ which in the sine-Gordon action provides the coordinate along the $O(2)$-invariant critical line of section~\ref{c1}. The amplitude $S_R$ vanishes and $S_T$ and $S_0$ become $-1$ when $\xi=\pi$, which then corresponds to the free fermion point $b^2=1/2$. The limit $\theta\to\infty$ (i.e. $m/\sqrt{s}\to 0$) of $S_T$, which is $e^{-i\frac{\pi}{2}(1+\frac{\pi}{\xi})}$, should coincide with the massless amplitude (\ref{S_O2}). Using (\ref{alpha}) and the condition at $\xi=\pi$ one obtains the relation
\EQ
\xi=\frac{\pi b^2}{1-b^2}\,,
\label{xi}
\EN
originally obtained from semiclassical and perturbative studies (\cite{ZZ} and references therein). The requirement that the energy density $\cos 2b\varphi$ is a relevant field implies $b^2<1$, so that $\xi\geq 0$. For $\xi<\pi$ the amplitudes $S_T$ and $S_R$ exhibit simple poles in the physical strip corresponding to soliton-antisoliton bound states $B_n$. The poles are located at $\theta=i(\pi-n\xi)$ and determine the masses of the particles $B_n$ as
\EQ
m_n=2m\sin\frac{n\xi}{2}\,,\hspace{1cm}1\leq n<\left[\frac\pi\xi\right]\,,
\label{mn}
\EN
where $[x]$ denotes the integer part of $x$. The lightest bound state $B_1$ is the particle created by the field $\varphi$; it is no longer in the spectrum of the theory for $\xi>\pi$, which is a regime of repulsive soliton-antisoliton interaction. The scattering amplitudes involving the particles $B_n$ can be obtained from those for $A$ and $\bar{A}$ exploiting bootstrap equations similar to that depicted in Fig.~\ref{fpu_bootstrap}; the poles of these new amplitudes show, in particular, that $B_n$ can be seen as the bound state on $n$ particles $B_1$ (see \cite{ZZ}). 

The fermionic form of the model corresponds to the scaling limit of two Ising models coupled through the product of their energy densities (the four-fermion term in (\ref{Thirring})). This is known as the Ashkin-Teller model and is defined on the lattice by the Hamiltonian
\EQ
{\cal H}_{AT}=-\sum_{\langle ij\rangle}\{J[\sigma_{1,i}\sigma_{1,j}+\sigma_{2,i}\sigma_{2,j}]+
J_4\sigma_{1,i}\sigma_{1,j}\sigma_{2,i}\sigma_{2,j}\}\,\,,
\label{AT}
\EN
where $\sigma_{1,i}$ and $\sigma_{2,i}$ are the two Ising spins at site $i$. This four-state model possesses a line of second order phase transition in the space $(J,J_4)$ which is described by the line of conformal field theories with $c=1$, and was originally cast in the bosonic/fermionic field theoretical framework in \cite{KB,Kadanoff}. For $J=J_4$ the Ising variables $\sigma_{1,i}$, $\sigma_{2,i}$ and $\sigma_{1,i}\sigma_{2,i}$ play a symmetric role in (\ref{AT}), corresponding to the four-state Potts model subspace. On the other hand, it follows from Table~\ref{table1} that the critical four-state Potts model has $c=1$ and $\Delta_\varepsilon=b^2=1/4$. As a consequence, the sine-Gordon scattering amplitudes with $\xi=\pi/3$ solve the scaling four-state Potts model; notice that $S_R$ vanishes at this point, so that we have purely transmissive scattering as in the three-state case, and one can check that the remaining amplitudes reduce to products of the blocks (\ref{fa}) (the same is true for all integer values of $\pi/\xi$). At the particle level the symmetry enhancement is signaled by the fact that for $\xi=\pi/3$ the mass of the particle $B_1$ becomes degenerate with that of $A$ and $\bar{A}$. More generally, the scattering description of the Ashkin-Teller model is given in \cite{AT,ATratios} and we will come back on this in section~\ref{wetting_section}.

\subsection{Antiferromagnets}
\label{antiferro}
An antiferromagnetic interaction corresponds to taking $J<0$ in a lattice Hamiltonian such as (\ref{ising}), so that configurations in which nearest neighboring spins take `opposite' values are energetically favored. Since the extent to which spins can differ from their neighbors depends on the number of neighbors, phase transitions in antiferromagnets are crucially dependent on the lattice structure and, in this sense, non-universal. Still, if an antiferromagnet on a given lattice undergoes a second order transition, this will be described by a field theory in the scaling limit. 

Considering for simplicity the square lattice, the Ising case is particularly simple, since it can be mapped back to the ferromagnetic case introducing the staggered spin variable $\Sigma_i=(-1)^{i_1+i_2}\sigma_i$, where $i=(i_1,i_2)$ is the site label. The three-state Potts model provides instead a non-trivial illustration of the combined effect of internal and lattice symmetries. Now there are infinitely many configurations in which each spin has a value different from that of its nearest neighbors, so that the model possesses infinite ground state degeneracy. These ground states admit an alternative representation in terms of arrows on the edges of the square lattice. After transferring the labels (colors) $s_i=1,2,3$ from sites to faces, an observer in a face with color $j$ looking across an edge to an adjacent face with color $j\pm 1$ (mod~3) puts an arrow on this edge pointing to his left/right. The arrow configurations obtained in this way satisfy at each site the rule ``two arrows in, two arrows out'' of the six-vertex model, an exactly solvable model \cite{LW,Baxter} which is critical and is described in the continuum by $c=1$ (Gaussian) conformal field theory. Hence the three-state Potts antiferromagnet on the square lattice is critical\footnote{More generally, the critical temperature of the antiferromagnetic $q$-state Potts model on the square lattice was determined in \cite{Baxter_AF}.} at $T=0$; this in turn means that its scaling limit for $T>0$ is described by the sine-Gordon model with energy density $\varepsilon\propto\cos 2b\varphi$, for some specific value of $b$. 

In order to further understand the nature of the correspondence with the Gaussian model, we consider the complex lattice spin variable $\sigma_j=e^{2i\pi s_j/3}$, and the staggered one $\Sigma_j=(-1)^{j_1+j_2}\sigma_j$, where $s_j=1,2,3$. Contrary to the energy density, $\Sigma_j$ and $\sigma_j$ carry a $Z_3$ charge associated to cyclic color permutations; $\Sigma_j$ is also odd under exchange of even and odd sublattices, namely the two sublattices identified by the parity of $j_1+j_2$. Hence in the scaling limit these variables correspond to charged scalar fields of the sine-Gordon model, in which charge is counted in units of the integer $m$ of (\ref{quantization}). Using the notation $U_m(x)=V_{m/4b}(z)\bar{V}_{-m/4b}(\bar{z})$ for the charged scalar fields, the natural identifications are $\Sigma=U_1$, $\sigma=U_{-2}=U^*_2$, consistent with the identification of the $Z_3$ charge with $m$~(mod~3), and of the sublattice parity with $(-1)^m$ \cite{AF}. For the conformal dimensions we have $\Delta_{\Sigma}=1/16b^2$, $\Delta_\sigma=1/4b^2$ and $\Delta_\varepsilon=b^2$; they match the lattice results (see \cite{CJS} and references therein) for $b^2=3/4$. The amplitudes (\ref{sg_tr}) with this value of $b^2$ yield the scattering solution of the scaling limit. As for the ferromagnetic case, there are two particles $A$ and $\bar{A}$, but for the antiferromagnet the fundamental charged field which creates $A$ is $\Sigma$ (rather than $\sigma$), and its odd sublattice parity forbids the fusion $AA\to\bar{A}$ characteristic of the ferromagnetic solution (\ref{3potts}).

The above symmetry identifications do not depend on $b$, and this suggests that different values of $b$ may describe the three-state Potts antiferromagnet on other lattices. It is known that at $T=0$ the model is critical also on the Kagom\'e lattice and allows for a Gaussian continuum limit \cite{HR}.

\subsection{Correlation functions}
\label{correlation}
The correlation functions are determined in the form (\ref{spectral}) if the form factors (\ref{ff}) are known. Remarkably, the form factors can also be determined when the $S$-matrix is known exactly. Indeed, in a disordered phase with factorized scattering the form factors of a scalar field local with respect to the order field satisfy the equations  \cite{KW,Smirnov}
\bea
&& F_n^\Phi(\theta_1,\ldots,\theta_i,\theta_{i+1},\ldots,\theta_n)=S(\theta_i-\theta_{i+1})F_n^\Phi(\theta_1,\ldots,\theta_{i+1},\theta_i,\ldots,\theta_n)\,,\label{ff1}\\
&& F_n^\Phi(\theta_1+2i\pi,\theta_2,\ldots,\theta_n)=F_n^\Phi(\theta_2,\ldots,\theta_n,\theta_1)\,,\label{ff2}\\
&& \mbox{Res}_{\theta'=\theta+i\pi}F_{n+2}^\Phi(\theta',\theta,\theta_1,\ldots,\theta_n)=i[1-\prod_{k=1}^nS(\theta-\theta_k)]F_n^\Phi(\theta_1,\ldots,\theta_n)\,,\label{ff3}
\eea
where we are considering considering the case of a theory with a single particle species (and then a single two-particle amplitude $S(\theta)$) which is sufficient to illustrate the general logic; we denote by $F_n^\Phi$ the matrix elements (\ref{ff}) as functions of rapidities. Eq.~(\ref{ff1}) simply states that in an integrable theory the exchange of the positions of two adjacent particles on the line produces a scattering amplitude. Eq.~(\ref{ff2}) is related to crossing and its meaning can be understood considering the case $n=2$. Then the equation can be written as $F_2^\Phi(\theta_1+i\pi,\theta_2)=F_2^\Phi(\theta_2,\theta_1-i\pi)$ and, considering that form factors of scalar fields depend on rapidity differences, says that $F_2^\Phi$ takes the same value at $i\pi+\theta$ and $i\pi-\theta$, with $\theta\equiv\theta_1-\theta_2$. Recalling Fig.~\ref{fpu_s_theta} this means that, contrary to the scattering amplitude, the form factor does not exhibit the crossing cut in the $s$-plane, but only the unitarity cut associated to Eq.~(\ref{ff1}). This is understandable, since the scattering amplitude has four legs and the form factor only two.

Eq.~(\ref{ff3}) prescribes the presence of a pole when charge, energy and momentum of two particles sum to zero. It is possible to set up a formal argument to see that these {\em annihilation} or {\em kinematical} poles are a peculiar two-dimensional implication of the general fact that matrix elements of $\Phi$ with $m$ particles on the left and $n$ particles on the right reduce, through crossing, to $F_{m+n}^\Phi$ plus ``disconnected parts'' due to annihilations \cite{Smirnov}. We will show in section~\ref{order} that kinematical poles are needed on physical grounds, since they account for phase separation in two dimensions \cite{DV_ps}. In theories exhibiting bound states form factors inherit from the scattering amplitudes the poles (\ref{pole}) and this allows one to write an additional residue equation relating $F_{n+1}^\Phi$ to $F_{n}^\Phi$. Kinematical and bound state poles are the only singularities of the form factors as functions of rapidities.

Here we illustrate the solution of the form factor equations for the Ising model ($S(\theta)=-1$), which minimizes the technicalities remaining non-trivial. In particular, above $T_c$ the order field $\sigma(x)$ creates the elementary excitations, so that spin reversal symmetry requires that $\sigma$ couples only to an odd number of particles. All the descendants of $\sigma$, however, share this property and the form factor equations (\ref{ff1}--\ref{ff3}) necessarily allow for infinitely many solutions for a fixed symmetry sector. It is a consequence of (\ref{shift}) that, among the scalar descendants, the fields $(\partial\bar{\partial})^k\phi$ have form factors behaving as $F_n^{(\partial\bar{\partial})^k\phi}\sim e^{k|\theta_i|}F_n^\phi$ as $|\theta_i|\to\infty$. It is then natural to expect, and can actually be shown more generally \cite{immf,D09}, that the primary field $\phi$ corresponds to the solution with mildest asymptotic behavior at high energies in the given symmetry sector. It is not difficult to check that for $\sigma$ such a solution is unique and reads \cite{BKW}
\EQ
F^\sigma_{2n+1}(\theta_1,\ldots,\theta_{2n+1})=i^n F^\sigma_1
\prod_{i<j}\tanh\frac{\theta_i-\theta_j}{2}\,,\hspace{1cm}T>T_c\,,
\label{sigmaodd}
\EN
where $F_1^\sigma$ is a constant by relativistic invariance.

When we move to $T<T_c$ we need to remember that, although $S$ is still $-1$, the excitations are kinks interpolating between the degenerate vacua $|0_+\rangle$ and $|0_-\rangle$. The field $\sigma$ is topologically neutral and couples to states starting and ending on the same vacuum, i.e. with an even number of kinks.
The fact that the kinks are non-local with respect to $\sigma$ (they are created by the disorder field $\mu$ of section~\ref{c2}) leads to a slight modification of Eqs.~(\ref{ff2}), (\ref{ff3}). Indeed, taking into account vacuum indices (\ref{ff2}) reads
\bea
\langle 0_+|\sigma(0)|K_{+-}(\theta_1+i\pi)K_{-+}(\theta_2)\ldots K_{-+}(\theta_{2j})\rangle &=& \nonumber \\
\langle 0_-|\sigma(0)|K_{-+}(\theta_2)\ldots K_{-+}(\theta_{2j})K_{+-}(\theta_1-i\pi)\rangle &=& \nonumber\\
- \langle 0_+|\sigma(0)|K_{+-}(\theta_2)\ldots K_{+-}(\theta_{2j})K_{-+}(\theta_1-i\pi)\rangle && \label{ff2kinks}
\eea
with the last equality following from spin reversal symmetry. Hence, if $F_n^\sigma$ denotes form factors on the vacuum $|0_+\rangle$, the r.h.s. of (\ref{ff2}) acquires a minus sign. Moreover, when $\theta_1=\theta_2$, the particle with rapidity $\theta_2$ can directly annihilate that with rapidity $\theta_1+i\pi$ in the first line of (\ref{ff2kinks}), while in the second line it has to scatter with $2j-2$ particles to reach and annihilate the particle with rapidity $\theta_1-i\pi$, thus producing a factor $S^{2j-2}$. These two different annihilation paths correspond to the two terms in the r.h.s. of (\ref{ff3}), so that the second one picks up the minus sign of the last line of (\ref{ff2kinks}). This is why the solution for $\sigma$ reads
\EQ
F^\sigma_{2n}(\theta_1,\ldots,\theta_{2n})=i^nF^\sigma_0\prod_{i<j}
\tanh\frac{\theta_i-\theta_j}{2}\,,\hspace{1cm}T<T_c\,,
\label{sigmaeven}
\EN
where $F_0^\sigma=\langle 0_+|\sigma|0_+\rangle$ is the spontaneous magnetization. The high energy limit, where the high and low temperature phases meet at the critical point, can also be used to show \cite{DSC,D09} that $F_2^\sigma(\theta_1,\theta_2)\to (F_1^\sigma)^2/F_0^\sigma$ as $\theta_1-\theta_2\to\infty$, a relation that fixes the relative normalization of form factors in the two phases. The energy density $\varepsilon$, on the other hand, is invariant under spin reversal and topologically neutral below $T_c$; it satisfies the same form factor equations above and below $T_c$ with solution $F^\varepsilon_n\propto\delta_{n,2}\,\sinh\frac{\theta_1-\theta_2}{2}$, consistently with the fact that it is a bilinear in the free fermions.

The form factors (\ref{sigmaodd}) and (\ref{sigmaeven}) determine $\langle\sigma(x)\sigma(0)\rangle$ in the Ising model through the large distance expansion (\ref{spectral}). For this case the expansion allows a resummation and the result can be exactly expressed in terms of a solution of a differential equation of Painlev\'e type, as originally shown in \cite{WMcTB} taking the scaling limit of the lattice result. For theories with interacting particles, however, resummations are not known, even if all form factors are available. It is then very relevant for practical purposes to see how effective partial sums of (\ref{spectral}) can be. Taking the example of the susceptibility $\chi=\int d^2x\langle\sigma(x)\sigma(0)\rangle$ in the Ising model, the leading contribution to the large distance expansion of the correlator comes from $F_1^\sigma$ above $T_c$ and from $F_2^\sigma$ below (the vacuum contribution is subtracted in the definition of the susceptibility). This two-particle approximation (i.e. taking into account contributions up to two particles) yields $\Gamma_+/\Gamma_-\simeq 37.699$ for the universal ratio of the susceptibility amplitudes defined in (\ref{chi}), to be compared with the exact result\footnote{See \cite{ising_ratios,review} for the complete list of exact amplitude ratios in the two-dimensional Ising universality class.} $\Gamma_+/\Gamma_-=37.6936..$ \cite{WMcTB}. The origin of such a remarkable agreement has to be understood as follows. The truncated form factor sum for a correlator $\langle\Phi(x)\Phi(0)\rangle$ becomes exact for $|x|$ large, but certainly fails to reproduce the exact divergence $|x|^{-4\Delta_\Phi}$ at short distance; the contribution of short distances, on the other hand, is suppressed upon integration over $d^2x=2\pi|x|d|x|$, and the suppression increases as $\Delta_\Phi$ decreases. The order field, in particular, has the smallest dimension and yields the best convergence for the integrated correlator. In this way the two-particle approximation becomes extremely effective in integrable theories, where two-particle form factors are not too difficult to compute, both for ordinary particles (see e.g. \cite{review,ATratios}; also \cite{Lukyanov} for a different approach) and for kinks \cite{DC,DCq4}. In Table~\ref{table2} we report the results of the two-particle approximation for $\Gamma_+/\Gamma_-$ in the $q$-state Potts model and compare them with the lattice results; for $q\neq 2$ the latter are obtained through series expansions or Monte Carlo simulations.

\begin{table}
\begin{center}
\begin{tabular}{|l|r|c|}
\hline
$q$           & \text{field theory (2pa)} & \text{lattice}  \\
\hline
$1$  &  160.2 \cite{DVC}  &    $162.5\pm 2$ \cite{JZ} \\

$2$  & 37.699 \cite{DC}    &  $37.6936..$  \cite{WMcTB} \\

$3$   & 13.85  \cite{DC}   &  13.83(8)  \cite{EG,SBB} \\

$4$   & 4.01  \cite{DC}     &  3.9(1) \cite{SJ} \\

\hline
\end{tabular}
\caption{Results for the universal susceptibility amplitude ratio $\Gamma_+/\Gamma_-$ in the $q$-state Potts model from field theory in the two-particle approximation (2pa) and from the lattice; $q=1$ corresponds to the amplitude ratio for the mean cluster size in percolation.}
\label{table2}
\end{center}
\end{table}

In the $q$-state Potts model above $T_c$ the probability that two sites have the same color $\alpha$ is given by
\EQ
\langle\delta_{s(x_1),\alpha}\delta_{s(x_2),\alpha}\rangle=\frac1q P_2(x_1,x_2)+\frac{1}{q^2}(1-P_2(x_1,x_2))\,,
\label{samecolor}
\EN
where $P_2$ is the probability that the two sites are in the same cluster. Recalling the definition (\ref{potts_spin}), we obtain 
\EQ
\langle\sigma_\alpha(x_1)\sigma_\alpha(x_2)\rangle=(q-1)P_2(x_1,x_2)\,,
\label{connectivity}
\EN
so that the susceptibility is proportional to the mean cluster size. When evaluating (\ref{samecolor}) below $T_c$ one has to take into account the possibility that the sites belong to the infinite cluster, but (\ref{connectivity}) is recovered in the limit $q\to 1$ relevant for percolation, with $P_2$ the probability that the sites are in the same finite cluster. Hence, with $\Delta_\sigma|_{q=1}=5/96$ and $\Delta_\varepsilon|_{q=1}=5/8$ from Table~\ref{table1}, we obtain that the mean size of finite clusters in percolation diverges with the susceptibility exponent $\gamma=(1-2\Delta_\sigma)/(1-\Delta_\varepsilon)|_{q=1}=43/18$ as $p\to p_c$. $(\Gamma_+/\Gamma_-)_{q=1}$, on the other hand, gives the universal amplitude ratio of the mean size of finite clusters on the two sides of the percolation transition; the agreement of the theoretical and numerical results for this number in Table~\ref{table2} has to be regarded as particularly remarkable since its value had been controversial for decades, with numerical estimates spanning different orders of magnitude (see \cite{PHA,Ziff}).

\section{Crossover behavior and massless scattering}
\label{crossover}
\subsection{Dilute Ising model}
\label{dilute}
We saw in section~\ref{fields} that for a given internal symmetry group $G$ there may exist multicritical points characterized by the presence of more than one $G$-invariant relevant field, so that the realization of such a multicriticality requires the tuning of more than one parameter. The basic example is obtained considering the Ising model and allowing for the presence of vacant sites, i.e. considering the Hamiltonian
\EQ
{\cal H}_\textrm{dilute\,Ising}=-J\sum_{\langle i,j\rangle}\sigma_i\sigma_jt_it_j+\Delta\sum_it_i\,,\hspace{1cm}\sigma_i=\pm 1\,,\,\,\,t_i=0,1\,,
\label{dilute_ising}
\EN
where $t_i=0$ introduce a vacancy at site $i$ and $\Delta$ controls the vacancy density; as $\Delta\to-\infty$ the configurations with vacancies are suppressed and the undilute case is recovered. Dilution preserves spin reversal symmetry, which is observed to break spontaneously along a line in the ($J^{-1},\Delta$)-plane, preserving the second order character of the undilute case up to a critical value of the vacancy density, and becoming first order above this value (Fig.~\ref{fpu_dilute}). The point on the line where the order of the transition changes provides the simplest example of a tricritical point, i.e. a scale invariant point where a second symmetry-invariant field, the dilution $\varepsilon'$ conjugated to $\Delta-\Delta_c$, is relevant together with the energy density $\varepsilon$ conjugated to $J-J_c$. Adding $\varepsilon'$ to the action of the tricritical point induces a shift along the phase transition line, the first or the second order portion depending on the sign of the coupling. The correlation length is infinite along the second order portion of the line, but there is no scale invariance, since $\varepsilon'$ is relevant and its conjugated coupling dimensionful; moving down along the line, correlators cross-over from the power law behavior ruled by the scaling dimensions of the tricritical point to that ruled by the scaling dimensions of the ordinary Ising critical point. We call crossover line the transition line connecting two scale invariant points.

\begin{figure}
\begin{center}
\includegraphics[width=6cm]{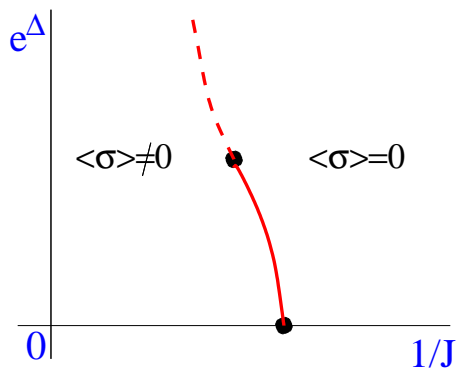}
\caption{Schematic phase diagram of the Ising model with vacancies; the pure model corresponds to $\Delta=-\infty$. The first order (dashed) and second order portions of the phase transition line meet at a tricritical point.}
\label{fpu_dilute}
\end{center} 
\end{figure}

In principle, the infinite correlation length allows for a continuous, field theoretic description along the whole crossover line and its vicinity. As we discussed in section~\ref{c2}, in two dimensions the Ising tricritical point corresponds to the case $t=4$ of the series of minimal conformal models; it has central charge $c=7/10$, $\varepsilon=\phi_{1,2}$ and $\varepsilon'=\phi_{1,3}$. Then the counting argument of section~\ref{integrability} says, in particular, that the transition line, corresponding to a shift in the $\phi_{1,3}$ direction, is integrable in the continuum. The crossover portion, which has infinite correlation length, must correspond to an integrable theory with massless elementary excitations. 

Massless particles are normally excluded from treatments relying on the analytic properties of the $S$-matrix \cite{ELOP}. The basic reason can be understood recalling Fig.~\ref{fpu_s_plane}, where as the mass $m$ tends to zero infinitely many branch points collapse in the origin, thus making the analytic structure untreatable. In an integrable theory, however, the scattering amplitude has only two cuts in the $s$-plane (Fig.~\ref{fpu_s_theta}) and the limit $m\to 0$ can be taken in a controlled way \cite{Alyosha}. Within the rapidity parameterization $p=m\sinh\beta$ of the momenta, one can take $m=Me^{-\alpha}$, $\alpha\to+\infty$, and obtain right movers with $e=p=(M/2)e^\theta$ for $\beta=\theta+\alpha$, and left movers with $e=-p=(M/2)e^{-\theta}$ for $\beta=\theta-\alpha$, with $M$ a mass scale. For the scattering of a right mover with rapidity $\theta_1$ and a left mover with rapidity $\theta_2$, the relativistic invariant (\ref{e2}) becomes
\EQ
s=M^2e^{\theta_1-\theta_2}\,,
\label{e2rl}
\EN
a relation which maps the branch point $s=0$ at $\theta_{12}\equiv\theta_1-\theta_2=-\infty$. This requires that the upper and lower edges of the unitarity cut, which in the massive case were mapped by (\ref{s_theta}) to real values of $\theta_{12}$ with opposite sign, now correspond to the boundary values of two different functions, $S(\theta_{12})$ for the upper edge and $\tilde{S}(\theta_{12})$ for the lower edge, with $\theta$ real; the upper and lower edges of the crossing cut ($s<0$) are obtained shifting $\theta_{12}$ by $i\pi$. With these notations and for a single particle species the massless limit of the unitarity and crossing equations (\ref{unitarity0}), (\ref{crossing0}) reads
\EQ
S(\theta)\tilde{S}(\theta)=1\,,\hspace{1cm}
S(\theta)=\tilde{S}(\theta+i\pi)\,,
\EN
which can be combined into
\EQ
S(\theta)S(\theta-i\pi)=1\,.
\label{crossunitarity1}
\EN

For the crossover line in the dilute Ising model, the mass scale $M$ appearing in (\ref{e2rl}) is associated to the coupling conjugated to the relevant field $\epsilon'$, which is proportional to $M^{2-2\Delta_{\varepsilon'}}$. The high energy limit $M^2/s\to 0$ in which this coupling becomes negligible is that towards the tricritical point, while the critical point of the undilute model is approached as $M^2/s\to\infty$. Since we know that the latter contains a single particle species, which is a free fermion, one looks for the simplest meromorphic solution of (\ref{crossunitarity1}) giving $-1$ when $\theta=-\infty$. This solution reads
\EQ
S(\theta)=\frac{e^\theta-i}{e^\theta+i}=\tanh\left(\frac{\theta}{2}-\frac{i\pi}{4}\right)\,,
\label{Stim}
\EN
and the thermodynamic Bethe ansatz can be used to confirm that it leads to the correct central charge $7/10$ in the high energy limit \cite{Alyosha}. The amplitude exhibits a pole at $\theta_0=-i\pi/2$ (corresponding to $s_0=-iM^2$) in the unphysical strip $\mbox{Im}\,\theta\in(-i\pi,0)$. According to the general interpretation of second sheet poles \cite{ELOP}, this corresponds to a resonance, i.e. to an unstable particle with lifetime proportional to $1/\sqrt{-\mbox{Im}\,s_0}=1/M$. As in the massive case, the scattering solution (\ref{Stim}) can be used to compute form factors and correlation functions \cite{DMS_flow}.

\subsection{Correlated percolation}
\label{correlated}
In previous sections we considered the case of {\it random} percolation, i.e. the percolation problem in which each site (or bond\footnote{Site and bond percolation belong the same universality class.}) is occupied with probability $p$ independently of the others. It is relevant, however, to consider also the case of {\it correlated} percolation, in which there is an interaction among the sites \cite{SA}. Once again the Ising model provides the basic example if we consider the {\it spin clusters} obtained connecting nearest neighboring sites with spin ``up", i.e. equal to $+1$. Indeed, let us consider the case in which a magnetic field $H$ is present, i.e. we take the Hamiltonian (\ref{ising}) and add the term $-H\sum_i\sigma_i$. Then for $H=+\infty$ a cluster of up spins covers the whole lattice, while no up spins are left at $H=-\infty$, indicating that a percolation transition takes place in between at some value $\hat{H}_c(1/\hat{J})$ (we introduce the notations $\hat{J}\equiv J/T$, $\hat{H}\equiv H/T$). We know that $\hat{H}_c(0)=0$, since at zero temperature the system is in the ferromagnetic ground state selected by the sign of $H$. For $J=0$, on the other hand, sites are uncorrelated and the transition takes place when the probability that a spin is up coincides with the random site percolation threshold $p_c^0$, i.e. $\hat{H}_c(\infty)$ is determined by the equation $e^{\hat{H}_c(\infty)}/(2\cosh\hat{H}_c(\infty))=p_c^0$.
It is natural to expect that $\hat{H}_c$ is a monotonic function of the interaction among the spins, and it can be shown \cite{CNPR} that the existence of a spontaneous magnetization implies the presence of an infinite cluster. Since two-dimensional lattices have $p_c^0\geq 1/2$ \cite{SA}, and then $\hat{H}_c(\infty)\geq 0$, these considerations lead to the phase diagram of Fig.~\ref{fpu_corr_perc}. While for $\hat{J}>\hat{J}_c$ the percolation transition is, like the magnetic one, of the first order, the points $(\hat{J}^{-1},\hat{H})=(\hat{J}_c^{-1},0)$ and $(\infty,\hat{H}_c(\infty))$ are fixed points for percolation (correlated and random, respectively). The second order transition line connecting them is the crossover line; the linear size of clusters diverges along this line\footnote{For the triangular lattice, which has $p_c^0=1/2$, and then $\hat{H}_c(\infty)=0$, the transition line remains at $H=0$ for any $J$.}, while the magnetic correlation length is everywhere finite away from the Curie point $(\hat{J}_c^{-1},0)$. This means that the field theory which describes percolation of spin clusters is not that which describes the local magnetization.

\begin{figure}
\begin{center}
\includegraphics[width=7cm]{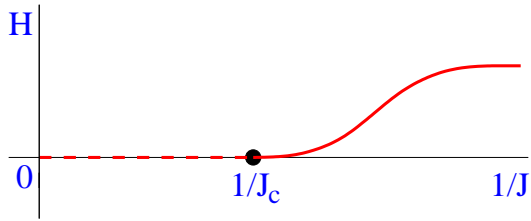}
\caption{Phase diagram for two-dimensional Ising percolation. An infinite cluster of up spins exists above the percolation line. The percolation transition changes from first order (dashed) to second order at the Curie point.}
\label{fpu_corr_perc}
\end{center} 
\end{figure}

In order to progress, let us recall what we saw for random percolation. In that case an auxiliary $q$-state Potts spin variable was associated to each lattice site, so that the ``geometric" percolation transition was mapped onto the thermal Potts transition, in the limit $q\to 1$ in which the auxiliary Potts degrees of freedom are removed. Exactly at $q=1$ there are no degrees of freedom, but at $q=1+\epsilon$, $\epsilon\ll 1$, we can extract the quantities relevant for percolation, e.g. the two-point connectivity which, recalling (\ref{connectivity}), is given by $P_2=\langle\sigma_\alpha\sigma_\alpha\rangle/\epsilon$. A similar procedure can be used to track the Ising spin clusters, associating the auxiliary $q$-state Potts variable $s_i$ to a site $i$ in which the Ising spin is up. In practice one adds to the Ising Hamiltonian the term $-J\sum_{\langle ij\rangle}t_it_j\delta_{s_i,s_j}$, where $t_i=(\sigma_i+1)/2=0,1$ \cite{Murata,CK}, and obtains in this way a dilute $q$-state Potts model, of which the dilute Ising model of the previous subsection is a particular case ($q=2$). As for $q=2$, the Potts model possesses, for real $q<4$, a tricritical point and a crossover  line connecting the tricritical point to the critical point of the undilute model \cite{NBRS}. For $q=1+\epsilon$ this line describes the second order percolation transition line of Fig.~\ref{fpu_corr_perc} \cite{CK}.

Moreover, also for $q\neq 2$ the field $\phi_{1,3}$ is responsible for the departure from the tricritical point towards the critical one \cite{Z_cth},  so that the whole transition surface spanned by $q$ is integrable, in particular the percolation transition line at $q=1+\epsilon$ \cite{isingperc}. The $S_q$-invariant massless scattering theory is not known for generic $q$, but the results of \cite{FSZ} are sufficient to show \cite{isingperc} that the amplitude $S_{RL}(\theta)$ possesses a resonance pole located at $\theta_0=-i\pi(t-3)/2$, where $t$ determines the central charge $c=1-6/[t(t+1)]$ along the tricritical line; $q=2$, with $c=7/10$, corresponds to $t=4$ and $\theta_0=-i\pi/2$, as we saw in the previous subsection. On the other hand, at $q=1$ the only degrees of freedom left are the vacancies, which are the original Ising variables, so that we have $c=1/2$ on the tricritical line, i.e. $t=3$ and $\theta_0=0$. The analytic mechanism behind the phase diagram of Fig.~\ref{fpu_corr_perc} is now clear. At $q=1+\epsilon$ the crossover line corresponds to the presence of an unstable particle with lifetime proportional to $1/\epsilon$, plus a number proportional to $\epsilon$ of massless particles (the Potts degrees of freedom); these massless particles yield an infinite connectivity length and account for the existence of the second order percolation line. Exactly at $q=1$ we are left with the theory which describes the local magnetization properties of the Ising model; there are no massless particles left, while the former unstable particle has become stable and accounts for the finite magnetic correlation length.

\section{Phase separation}
\label{phase_separation}
\subsection{Order parameter profile}
\label{order}
So far we considered systems defined on the whole plane. In this section we start the study of effects produced by the presence of boundaries, beginning with the case of an infinitely long strip of with $R$. For discrete internal symmetry and $T<T_c$, a boundary field on the two edges of the strip can be used to drive the system in one of the degenerate phases as $R\to\infty$. We have already seen in section~\ref{away} that these phases are in correspondence with the degenerate vacua $|0_a\rangle$, $a=1,\ldots,n$, of the associated quantum theory, and that the kinks $K_{ab}(\theta)$ provide the elementary excitations. On the other hand, if we place the strip as in Fig.~\ref{fpu_strip}, and apply on both edges a boundary field favoring a phase $a$ for $x<0$ and one favoring a different phase $b$ for $x>0$, for $R$ large we will have different pure phases at the far left and far right separated by a central {\it interfacial region}; $R$ large means much larger than the correlation length $\xi$ of the pure phases. 
Field theory provides a general and exact description of such a phase separation in the scaling limit \cite{DV_ps,DS_double}.

\begin{figure}
\begin{center}
\includegraphics[width=9cm]{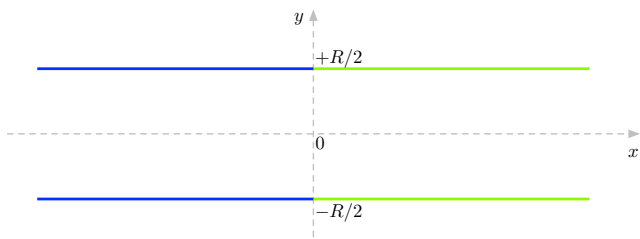}
\caption{Strip geometry.}
\label{fpu_strip}
\end{center} 
\end{figure}

To see this we observe first of all that within the quantum picture with the vertical coordinate $y$ as imaginary time, the boundary conditions play the role of initial and final states and can be expanded over the complete basis of asymptotic states of the bulk theory. In particular, the mixed boundary conditions specified above (we will refer to them as $ab$ boundary conditions) correspond to\footnote{In general, when the change from boundary condition $a$ to boundary condition $b$ takes place at $x_0\neq 0$, (\ref{Bab}) exhibits the factor $e^{-iPx_0}$.}
\EQ
|B_{ab}(\pm R/2)\rangle=e^{\pm\frac{R}{2}H}\left[\int\frac{d\theta}{2\pi}f(\theta)|K_{ab}(\theta)\rangle+\ldots\right]\,,
\label{Bab} 
\EN
where we are first considering the basic case in which the vacua $|0_a\rangle$ and $|0_b\rangle$ are {\it adjacent}, i.e. there exists a single-kink excitation $|K_{ab}(\theta)\rangle$ connecting them, and the dots stay for states with larger total mass yielding subleading corrections in the large $R$ limit we are interested in. Then the partition function for the strip with $ab$ boundary conditions can be written as
\EQ
Z_{ab}(R)=\langle B_{ab}(R/2)|B_{ab}(-R/2)\rangle\sim\int\frac{d\theta}{2\pi}|f(\theta)|^2e^{-m_{ab}R\cosh\theta}\sim\frac{|f(0)|^2}{\sqrt{2\pi m_{ab}R}}\,e^{-m_{ab}R}\,,
\label{Zab}
\EN
where $m_{ab}$ is the kink mass and we took the large $R$ limit, as  we will regularly do below\footnote{Also for later use we recall the formula $\int_{-\infty}^\infty dx\,e^{-px^2+2qx}=e^{q^2/p}\sqrt{\pi/p}$.}. For uniform boundary conditions of type $a$ the expansion of the boundary state $|B_a\rangle$ starts with the contribution of the vacuum $|0_a\rangle$, and the corresponding partition function reads $Z_{a}(R)=\langle B_{a}(\frac{R}{2})|B_{a}(-\frac{R}{2})\rangle\sim\langle 0_a|0_a\rangle=1$. Then we can determine the {\it interfacial tension}
\EQ
\Sigma_{ab}=-\lim_{R\to\infty}\frac{1}{R}\,\ln\frac{Z_{ab}(R)}{Z_a(R)}=m_{ab}\,.
\label{excess}
\EN

We now use the notations $\sigma(x,y)$ for the order field at the point $(x,y)$, and $\langle\sigma\rangle_a\equiv\langle 0_a|\sigma(x,y)|0_a\rangle$ for the order parameter in the pure phase $a$. The order parameter profile in the middle of the strip is
\bea
\langle\sigma(x,0)\rangle_{ab} &=& \frac{1}{Z_{ab}}\langle B_{ab}(R/2)|\sigma(x,0)|B_{ab}(-R/2)\rangle
\nonumber\\
&\sim & \frac{|f(0)|^2}{Z_{ab}}\int\frac{d\theta_1}{2\pi}\frac{d\theta_2}{2\pi}\langle K_{ab}(\theta_1)|\sigma(0,0)|K_{ab}(\theta_2)\rangle e^{-m[(1+\frac{\theta_1^2}{4}+\frac{\theta_2^2}{4})R-i\theta_{12}x]}\,,
\label{opp1}
\eea
where $\theta_{12}\equiv\theta_1-\theta_2$ and for simplicity we write $m$ instead of $m_{ab}$. The matrix element inside the integral contains a disconnected term $2\pi\delta(\theta_{12})\langle\sigma\rangle_b$ (corresponding to annihilation of the two particles) which contributes an additive constant to the profile; due to crossing, instead, the connected part is the two-particle form factor $\langle 0_b|\sigma(0,0)|K_{ba}(\theta_1+i\pi)K_{ab}(\theta_2)\rangle$. For the Ising model, recalling (\ref{sigmaeven}), this is given by $F_2^\sigma(\theta_1+i\pi,\theta_2)$, and exhibits the annihilation pole at $\theta_1=\theta_2$ with residue $2i\langle\sigma\rangle_+=i[\langle\sigma\rangle_+-\langle\sigma\rangle_-]$. This pole and its residue owe nothing to the integrability of the Ising model. Indeed, integrability simplifies the scattering, but the matrix element in (\ref{opp1}) is taken between states involving a single particle, and then no scattering. The pole then is always present as a consequence of the non-locality of the kinks with respect to the order parameter, and for a generic model the residue is $i[\langle\sigma\rangle_a-\langle\sigma\rangle_b]$ \cite{DC}. We write
\EQ
\langle K_{ab}(\theta_1)|\sigma(0,0)|K_{ab}(\theta_2)\rangle_\textrm{connected}=i\frac{\langle\sigma\rangle_a-\langle\sigma\rangle_b}{\theta_{12}}+\sum_{n=0}^\infty c_n\,\theta_{12}^n\,,
\label{F2conn}
\EN
take the derivative of (\ref{opp1}) with respect to $x$ in order to get rid of the pole, perform the integrations over rapidities, integrate back over $x$ with the boundary condition $\langle\sigma(+\infty,0)\rangle_{ab}=\langle\sigma\rangle_b$, and obtain\footnote{We take into account that, for phases  $a$ and $b$ playing a symmetric role, $f(\theta)=f(0)+O(\theta^2)$.} \cite{DV_ps}
\EQ
{\langle\sigma(x,0)\rangle_{ab}=\frac{1}{2}\bigl[\langle\sigma\rangle_{a}+\langle\sigma\rangle_b\bigr]-\frac{1}{2}\bigl[\langle\sigma\rangle_a-\langle\sigma\rangle_b\bigr]\,\mbox{erf}\Bigl(\sqrt{\frac{2m}{R}}\,x\Bigr)}{+c_0\sqrt{\frac{2}{\pi mR}}\,e^{-2mx^2/R}+\ldots}\,,
\label{opp}
\EN
where the error function $\mbox{erf}(z)\equiv\frac{2}{\sqrt{\pi}}\int_{0}^{z}dt\,e^{-t^2}$ interpolates between $-1$ and $1$, so that the profile interpolates between $\langle\sigma\rangle_a$ and $\langle\sigma\rangle_b$. Hence, the pole in (\ref{F2conn}) is responsible for phase separation in two dimensions while, as we are going to see, the regular terms characterize the interface structure. For the Ising model ($\langle\sigma\rangle_+=-\langle\sigma\rangle_-$, $c_0=0$) (\ref{opp}) reduces to $\langle\sigma\rangle_{-+}\sim\langle\sigma\rangle_+\,\mbox{erf}(\sqrt{\frac{2m}{R}}x)$, a result originally obtained from the lattice in \cite{Abraham_81} (see also \cite{Abraham} and references therein).

\subsection{Interfaces}
\label{interfaces}
The notion of interface emerges from (\ref{opp}) in the following way. In first approximation we think of it as a simple curve connecting the boundary condition changing points $(0,\pm R/2)$, intersecting the axis $y=0$ only once and sharply separating two pure phases (Fig.~\ref{fpu_interfaces}a); then we add corrections to this picture allowing for a structure localized on the curve.  Hence, we write as
\EQ
\sigma_{ab}(x|u)=\Theta(u-x)\langle\sigma\rangle_a+\Theta(x-u)\langle\sigma\rangle_b+A_0\delta(x-u)+A_1\delta'(x-u)+\ldots
\EN
the order parameter at the point $(x,0)$ in a configuration in which the curve intersects the central axis at $(u,0)$; $\Theta(x)$ is the step function equal to 1 if $x>0$ and to $0$ if $x<0$, while the delta functions account for the structure on the curve. Calling $p(u)du$ the probability that the curve passes in the interval $(u,u+du)$ on the axis $y=0$, we then have
\EQ
\langle \sigma(x,0) \rangle_{ab}= \int d u \, \sigma_{ab}(x \vert u) \, p(u)\,,
\EN
which coincides with the field theoretical result (\ref{opp}) for $p(x)=\sqrt{\frac{2m}{\pi R}}\,e^{-2mx^2/R}$ and $A_0=\frac{c_0}{m}$. We see that the term proportional to $c_0$ in (\ref{opp})  accounts for the first deviation from sharp phase separation, an effect which is natural to ascribe to bifurcation and recombination of the interface (Fig.~\ref{fpu_interfaces}b); it requires three different phases, consistently with the fact that $c_0=0$ in the Ising model. For the latter, the leading correction will come from trifurcations (Fig.~\ref{fpu_interfaces}c). In the three-state Potts model, on the other hand, $\langle\sigma_3\rangle_1=\langle\sigma_3\rangle_2$ and $c_0\neq 0$ \cite{DV_ps}, so that the leading variation in the profile $\langle\sigma_3(x,0)\rangle_{12}$ is produced by the Gaussian term in (\ref{opp}), corresponding to small drops of phase 3 forming on the interface between phases 1 and 2; the effect is suppressed as $R^{-1/2}$ as $R\to\infty$.

\begin{figure}
\begin{center}
\includegraphics[width=7cm]{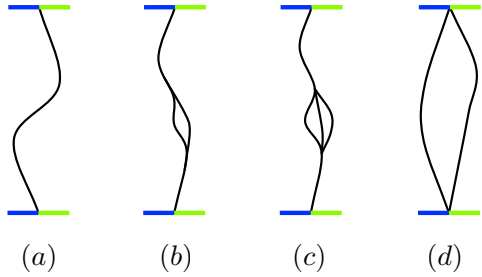}
\caption{Configurations of a single interface (a,b,c) and a double interface (d).}
\label{fpu_interfaces}
\end{center} 
\end{figure}

Using (\ref{shift}) the order parameter calculations that we performed for $y=0$ can be repeated for any fixed $|y|<R/2$, in the limit $R\to\infty$ \cite{DS_double}. From the order parameter one then obtains the probability density 
\EQ
p(x;y)=\frac{1}{\kappa}\sqrt{\frac{2m}{\pi R}}\,e^{-\chi^2}
\label{passage_strip}
\EN
for the passage in the interval $(x,x+dx)$ on a line of constant $y$; we introduced the definitions 
\EQ
\kappa\equiv\sqrt{1-4y^2/R^2}\,,\hspace{1cm}\chi\equiv\sqrt{\frac{2m}{R}}\,\frac{x}{\kappa}\,,
\label{kappa_chi}
\EN
so that (\ref{passage_strip}) exhibits a $y$-dependent variance which shrinks to zero as $y$ approaches the boundary condition changing points, as it should.

The above results have been obtained  for adjacent vacua $|0_a\rangle$ and $|0_b\rangle$. In some cases, however, connecting the two vacua requires a $n$-kink excitation with $n>1$. The $q$-state Potts model can be used to illustrate this case. For $q\leq 4$ the transition is of the second  order and below $T_c$ there are $q$ degenerate vacua which, by symmetry, are adjacent to each other (Fig.~\ref{fpu_adjacency}a). In presence of vacancies for $q<4$ there is also a tricritical point and a first order transition line (Fig.~\ref{fpu_dilute}) on which the ferromagnetic vacua $|0_i\rangle $, $i=1,\ldots,q$, are degenerate with the disordered vacuum $|0_0\rangle$. In this case the kinks $K_{i0}$ and $K_{0i}$ are the elementary ones, since two ferromagnetic vacua can be connected by a two-kink excitation $K_{i0}K_{0j}$. This first order transition is integrable in the scaling limit and the exact $S$-matrix \cite{D_dilute} shows that $K_{i0}K_{0j}$ do not form bound states, so that the ferromagnetic vacua are not adjacent to each other (Fig.~\ref{fpu_adjacency}b). The same conclusion holds for the pure model with $T=T_c$ and $q=5$, which also is a first order transition point. Strictly speaking, the scaling limit for $q>4$ is only defined in the limit $q\to 4^+$ in which the correlation length tends to infinity, but for $q=5$ (and actually up to $q\approx 10$) the correlation length is much larger than lattice spacing and an effective $S$-matrix description is possible and accurate \cite{DCq4}. 

When  the lightest excitation connecting $|0_a\rangle$ and $|0_b\rangle$ is a two-kink one, the boundary states (\ref{Bab}) become
\EQ
|B_{ab}(\pm R/2)\rangle=e^{\pm\frac{R}{2}H}\left[\sum_c\int d\theta_1 d\theta_2\,f_{acb}(\theta_1,\theta_2)\,|K_{ac}(\theta_1)K_{cb}(\theta_2)\rangle+\ldots\right]\,,
\EN
and give rise to the double interface of Fig.~\ref{fpu_interfaces}d. The calculation of the order parameter profile starting from these boundary states is a generalization of that performed above for adjacent phases \cite{DS_double}. For the Potts first order transition points discussed a moment ago one obtains, in particular,
\EQ
\langle\sigma_3(x,0)\rangle_{12}\sim\frac{\langle\sigma_3\rangle_1}{2}\left(1-\frac{2}{\pi}\,e^{-2z^2}-\frac{2z}{\sqrt{\pi}}\,\mbox{erf}(z)e^{-z^2}+\mbox{erf}^2(z)\right)\,,\hspace{1cm}z\equiv\sqrt{\frac{2m}{R}}\,x
\EN
as the leading term for $mR\gg 1$. Now $\lim_{R\to\infty}\langle\sigma_3(0,0)\rangle_{12}\neq\langle\sigma_3\rangle_1$, in contrast to what we saw for the pure Potts below $T_c$; the difference is that now a macroscopic drop of the disordered phase forms between the two interfaces. 

\begin{figure}
\begin{center}
\includegraphics[width=10cm]{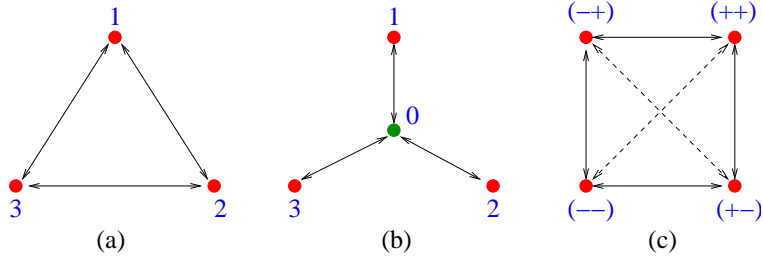}
\caption{Vacuum and kink structure for the three-state Potts model below $T_c$ (a), the dilute three-state Potts model at the first order transition (b), and the Ashkin-Teller model below $T_c$; in the latter case the dashed lines correspond to kinks present only in the attractive regime.}
\label{fpu_adjacency}
\end{center} 
\end{figure}

As for the single interface, the results for the order parameter profile can be extended to $y\neq 0$ and used to obtain the probability density that the two interfaces pass at points $x_1$ and $x_2$ on a line of constant $y$ \cite{DS_double}. This reads
\EQ
p(x_1,x_2;y)=\frac{2m}{\pi\kappa^2 R}\,(\chi_1-\chi_2)^2\,e^{-(\chi_1^2+\chi_2^2)}\,;
\label{double_passage}
\EN
the quadratic prefactor prevents the interfaces from passing at the same point and provides the effective repulsion which inflates the bubble of the third phase. In the field theory derivation it arises quite generally from the fermionic low-energy behavior of the scattering amplitudes \cite{DS_double}, and this explains the appearance of (\ref{double_passage}) in the framework of the lattice Ising model, suitably reinterpreted to account for the lack of a third phase \cite{AU,KoF}; (\ref{double_passage}) is also at the basis of the heuristic treatments of mutually avoiding interfaces \cite{Fisher_vicious}.

Let us remark, before closing this subsection, that the interfaces we are concerned with here and below are associated to separation of coexisting ordered phases and that, as we have shown, their properties can be derived directly in the continuum from the determination of the order parameter. The situation is different if the system is exactly at a second order phase transition point, where there are no coexisting phases and $\langle\sigma(x,0)\rangle_{ab}$ goes to zero for any $x$ as $R\to\infty$. Still, if we consider the Ising model at $T_c$ on the triangular lattice with $+-$ boundary conditions, any configuration exhibits a path (on the dual honeycomb lattice) with positive spins on its left and negative spins on its right connecting the boundary condition changing points $(0,\pm R/2)$. It has been shown in recent years that the passage probability for such a path can be determined exactly within the mathematical theory of stochastic curves subject to the constraint of conformal invariance \cite{Schramm}. In conformal field theory the same result emerges \cite{BB} as the solution of the second order differential equation associated to the degenerate field which induces the change of boundary conditions at the points $(0,\pm R/2)$ (see e.g. \cite{Cardy_sle,BBreport} for reviews). When passing from Ising to the three-state Potts model, the paths connecting the boundary condition changing points can branch even on the honeycomb lattice, and this case has been studied at criticality using Monte Carlo \cite{GC} and transfer matrix \cite{DJS} techniques. Looking for a relation between these lattice paths at criticality and the off-critical interfaces characterized earlier in this section directly in the continuum appears as an interesting subject for future research.

\subsection{Wetting}
\label{wetting_section}
We have seen how different vacuum connectivities give rise to two different regimes of phase separation:  a single interface with small bubbles along it (adjacent vacua) or a double interface enclosing a macroscopic layer of a third phase (non-adjacent vacua). The change of a parameter of the system may induce a transition from the first to the second regime which, thinking of liquid droplets which condensate to form a layer, is known as {\it wetting} transition (see e.g. \cite{Dietrich}). 

The Ashkin-Teller model provides an exact illustration of such a wetting transition \cite{DS_double}. The lattice Hamiltonian (\ref{AT}) shows that the model corresponds to two Ising models coupled through the product of their energy densities, and we have seen that this gives rise to a line of critical points in the space of the couplings $J$ and $J_4$, and that the scaling region around this line is described by the sine-Gordon model. In the regime in which the reversal symmetries of both Ising spins are spontaneously broken the model possesses four degenerate ferromagnetic vacua $|0_{a_1,a_2}\rangle$, $a_j=\pm$. The fundamental kinks interpolating between them are associated to the soliton $A=A_1+iA_2$ and antisoliton $\bar{A}=A_1-iA_2$ of sine-Gordon, since $A_j$ corresponds to the kink of the $j$-th Ising model. Sine-Gordon possesses the parameter (\ref{xi}) which serves as the coordinate along the critical line, with the free fermion point $\xi=\pi$ corresponding to the Ashkin-Teller decoupling point $J_4=0$. In the sine-Gordon repulsive regime $\xi>\pi$ ($J_4<0$) $A_1$ and $A_2$ are the only kinks and connect the vacua along the sides of the square in Fig.~\ref{fpu_adjacency}c;  vacua separated by the diagonal are not-adjacent since connecting them requires a two-kink state $A_1A_2$. In the attractive regime $\xi<\pi$ ($J_4>0$), instead, $A_1A_2$ form bound states and all vacua become adjacent. In particular, if we consider the strip with boundary conditions $--$ for $x<0$ and $++$ for $x>0$, we have a single interface with interfacial tension $\Sigma_{--,++}=m_1=2m\sin(\xi/2)$ (the lightest bound state mass given by (\ref{mn})) in the attractive regime, and a double interface (enclosing a mixture of the phases $+-$ and $-+$) with interfacial tension $2m$ in the repulsive regime. There is then a wetting transition at $J_4=0$ corresponding the unbinding of the bound state ($m_1=2m$).

A different mechanism, {\it kink confinement} \cite{conf1,conf2}, is responsible for the change of the interfacial structure when we start from a first order transition point in which ordered phases coexist with the disordered one (we gave above the examples of the dilute ($q<4$) and pure ($q>4$) Potts model), and then move the temperature below $T_c$. Let us first discuss kink confinement independently from wetting, within the simplest context of the Ising model. If we start at $T<T_c$ and add to the action a symmetry breaking term $h\int d^2x\,\sigma(x)$, with $h$ small, this removes the degeneracy of the two vacua inducing a difference $\delta{\cal E}\sim 2h\langle\sigma\rangle_+$ in their energy densities. An immediate consequence is that the kink $K_{+-}$ is no longer a stable particle. Technically this appears through the fact that for small $h$ the correction to the kink mass is given by $h\langle K_{+-}(\theta)|\sigma(0)|K_{-+}(\theta)\rangle=hF_2^\sigma(\theta+i\pi,\theta)$, which is infinite due to the annihilation pole in (\ref{sigmaeven}); as expected, explicit symmetry breaking removes topologically charged states from the spectrum and, as a consequence, forbids phase separation at equilibrium. On the other hand, if the sign of $h$ is such to favor $|0_+\rangle$ as the new ground state, the fate of the topologically neutral composite $K_{+-}K_{-+}$ is different. Indeed, to first order in $h$ the energy of a configuration in which the two particles are separated by a distance $R$ on the line grows with $R$ as $R\,\delta{\cal E}$. This is a confining potential which prevents the kinks from moving infinitely far apart and gives rise to a string of ``mesons" (kink-antikink composites). The lightest of these correspond to confinement of slow kinks and their masses $M_n$ are determined by the Schrodinger equation with the linear potential as
\EQ
M_n\sim 2m+\frac{(\delta{\cal E})^{2/3}z_n}{m^{1/3}}\,,
\label{mesons}
\EN
where $z_n$, $n=1,2,\ldots$, are positive numbers determined by the zeros of the Airy function, $\mbox{Ai}(-z_n)=0$. The spectrum (\ref{mesons}) was first obtained from the study of the spin-spin correlation function on the lattice \cite{McW}; the evolution of the mesonic spectrum as the magnetic field becomes stronger can also be followed \cite{conf1,FZspectroscopy,DGM}. In the three-state Potts model a symmetry breaking field favoring the vacuum $|0_1\rangle$ leads also to the formation of ``baryons" from the confinement of three-kink states $K_{12}K_{23}K_{31}$ \cite{DGconf,LTD,Rutkevich}.

Going back to the $q$-state Potts model at a point of coexistence of the ferromagnetic phases $i=1,\ldots,q$ with the disordered phase $0$, we know from Fig.~\ref{fpu_adjacency}b that boundary conditions on the strip of type $i$ for $x<0$ and $j$ for $x>0$ lead to the formation of a wetting intermediate disordered phase produced by the state $K_{i0}K_{0j}$. When moving the temperature slightly below $T_c$, the energy density of the disordered vacuum exceeds that of the ferromagnetic vacua by $\delta{\cal E}\propto(T_c-T)$, and the state $K_{i0}K_{0j}$ gives rise to stable ``mesonic kinks'' $\tilde{K}_{ij}^{(n)}$ with the masses (\ref{mesons}). Thus we go back to the connectivity depicted in Fig.~\ref{fpu_adjacency}a, i.e. the phases $i$ and $j$ are separated by a single interface.

\subsection{Quenched disorder}
\label{quenched}
The results seen so far allow few considerations about quenched disorder which, in a slightly different form, were originally developed in \cite{Kardar}. Ferromagnetic quenched disorder corresponds to replacing the constant coupling $J$ among sites in the lattice Hamiltonian with site-dependent couplings $J_{ij}$ randomly distributed (e.g. in a Gaussian way) around a positive mean; one then write the partition function $Z(\{J_{ij}\})$ summing over spin configurations for a given realization of the disorder, and takes as free energy of the system the average over the disorder of $-\ln Z$. A usual way to deal with such an average is to write\footnote{We denote by $\overline{X}$ the average of $X$ over the disorder.}
\EQ
\overline{\ln Z}=\lim_{n\to 0}\frac{\overline{Z^n}-1}{n}\,,
\label{replica}
\EN
so that one is led to consider the partition function $Z^n$ of $n$ copies (replicas) of the system with the same bond configuration. For weak disorder one can then argue that the average results into an effective interaction among $n$ replicas of the pure system ($J_{ij}=J$) (see e.g. \cite{Cardy_book}). Apart from the averaged free energy, the quantities $\overline{Z^n}=\overline{e^{n\ln Z}}$ determine the fluctuations of the random variable $\ln Z$. 

Consider now an Ising model on the strip geometry of Fig.~\ref{fpu_strip}, with boundary conditions $-$ for $x<0$ and $+$ for $x>0$, in presence of weak quenched disorder. Following the replica treatment, for temperatures below the critical temperature of the pure system and for $R$ much larger than the correlation length, we find $n$ interfaces connecting the boundary condition changing points $(0,\pm R/2)$. Assuming that, as indicated by perturbative calculations \cite{DD}, the phase transition persists in the disordered system, we have to think that the interaction among the $n$ interfaces induced by the disorder is attractive and binds them into the single interface required for a system with two degenerate vacua; for weak disorder the binding will be weak. It is well known (see e.g. \cite{Zamo_spectroscopyII} and references therein), that in $1+1$ dimensions the binding of $n$ identical particles becomes model-independent for weak attraction, since it approaches the non-relativistic limit of pairwise delta function interaction. Hence, in particular, we can read off the result from the sine-Gordon model of section~\ref{generic}, where we saw that the particles $B_n$, with masses $m_n$ given by (\ref{mn}), are bound states of $n$ particles $B_1$. In the weak binding limit $\xi\to 0$ we obtain $m_n\sim M(n-n^3\xi^2/24)$ for the interfacial tension in the disordered system. Recalling (\ref{Zab}) we then obtain $\overline{Z_{-+}^n}\propto(m_nR)^{-1/2}e^{-m_nR}$ for $R$ large. The fact that the leading effect of the disorder comes with the term $n^3$ in the exponent is a relevant result for two-dimensional systems with quenched disorder, since it can be shown that it leads to characteristic fluctuations of $\ln Z_{-+}$ of the form $R^{1/3}$ (rather than $R^{1/2}$) at large $R$ \cite{Kardar}.

\section{Interfaces at boundaries}
\label{interfacesatboundaries}
\subsection{Boundary scattering}
\label{boundary}
Critical and near-critical phenomena at boundaries provide an important chapter of the theory of universal behavior in statistical physics (see e.g. \cite{Binder,Diehl}). In two dimensions, the critical case is described by boundary conformal field theory \cite{Cardy_bcft}, which enables one to obtain exact results for boundaries of any shape exploiting conformal mappings and infinite dimensional conformal symmetry. Here we are interested in the near-critical case, and in this subsection consider the interaction of the bulk degrees of freedom with the boundary, for the basic case of the half-plane with coordinates $(x\geq 0,y)$ and translation invariant boundary conditions. Within the scattering framework in which $y$ is the imaginary time, this corresponds to consider particles moving on the half-line $x>0$; they scatter among them but also with the boundary, which one can conveniently think as an infinitely massive particle stuck at $x=0$. It is possible to follow a logic similar to that used for the scattering in the whole two-dimensional space-time and to look for exact solutions characterized by complete elasticity and factorization of the scattering amplitudes \cite{GZ} (see also \cite{FringK}). As for the whole plane, these solutions exist and describe the main interesting models; they also explicitly exhibit features, such as boundary bound states, which are general and will be recalled at some points later.

\begin{figure}
\begin{center}
\includegraphics[width=7cm]{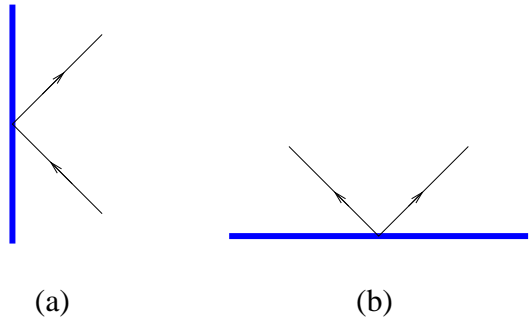}
\caption{Reflection in the direct channel (a) is related to pair emission in the crossed channel (b).}
\label{fpu_boundary}
\end{center} 
\end{figure}

Integrability on the half-plane requires that on the whole plane to start with, and two-particle scattering is specified by a known amplitude $S(\theta_1-\theta_2)$ of the bulk theory; we are referring to the case of a single particle species which is sufficient to illustrate the main ideas. Due to complete elasticity and translation invariance in the time direction, a particle traveling towards the boundary with rapidity $\theta<0$ will bounce back with rapidity $-\theta$ (Fig.~\ref{fpu_boundary}a), and we denote by $R(\theta)$ the amplitude for such a reflection process. Recalling that the boundary can be seen as a particle with zero rapidity, the unitarity equation
\EQ
R(\theta)R(-\theta)=1
\label{boundary_unitarity}
\EN
for the reflection amplitude is a particular case of (\ref{unitarity}). The boundary version of crossing symmetry, on the other hand, is more subtle and can be understood as follows \cite{GZ}. If we look at the process of Fig.~\ref{fpu_boundary}a from the ``crossed channel'', keeping imaginary time along the $y$ direction, it will appear as in Fig.~\ref{fpu_boundary}b, i.e. as the emission from the boundary of a pair of particles with rapidities $\theta$ and $-\theta$ (momentum, rather than energy, is conserved in the crossed channel). Hence, the reflection amplitude in the direct channel is related by analytic continuation to the pair emission amplitude, that we denote $K(\theta)$, in the crossed channel. Since in passing from Fig.~\ref{fpu_boundary}a to Fig.~\ref{fpu_boundary}b energy and momentum have interchanged their role, the analytic continuation following from (\ref{rapidity}) is
\EQ
K(\theta)=R(\frac{i\pi}{2}-\theta)\,.
\label{RK}
\EN
On the other hand, interchanging the position of the particles with rapidities $\theta$ and $-\theta$ in the crossed channel produces a factor $S(2\theta)$; then we have the relation
\EQ
K(\theta)=S(2\theta)K(-\theta)\,,
\label{K_unitarity}
\EN
which through (\ref{RK}) provides a second equation for $R(\theta)$ \cite{GZ}. 

For the Ising model, with $S(\theta)=-1$, the equations (\ref{boundary_unitarity}-\ref{K_unitarity}) admit the one-parameter solution
\EQ
R_h(\theta)=i\tanh\left(\frac{i\pi}{4}-\frac{\theta}{2}\right)\,\frac{\kappa-i\sinh\theta}{\kappa+i\sinh\theta}\,,
\label{R_ising}
\EN
corresponding, for $T<T_c$, to the presence of a boundary magnetic field $h$ related to $\kappa$ as \cite{GZ}
\EQ
\kappa=1-\frac{h^2}{2m}\,.
\label{kappa_h}
\EN
In this case the identification of the boundary condition is allowed by the fact that the boundary operator conjugated to $h$ has dimension $\Delta=1/2$ \cite{Cardy_89}, and then is linearly expressible in terms of the free fermions of the bulk theory; this allows one to obtain (\ref{R_ising}) and (\ref{kappa_h}) through free field techniques \cite{GZ}. Notice that $R_0(\theta)=-i\coth(\frac{i\pi}{4}-\frac{\theta}{2})$ possesses a pole at $\theta=i\pi/2$ which is absent for $h\neq 0$ and turns out to be a general feature distinguishing symmetry-preserving (or free) from symmetry-breaking boundary conditions. Together with minimality this normally allows one to identify solutions of the boundary scattering problem corresponding to fixed ($h=\infty$) and free boundary conditions below $T_c$ (see e.g. \cite{Chim} for the Potts model). Identifications below $T_c$ are made easier by the fact that the bulk excitations are kinks whose trajectory divides the half-plane in Fig.~\ref{fpu_boundary} into magnetized domains which have to be compatible with the magnetization state of the boundary. Information about the system above $T_c$ can usually be obtained by duality considerations (see \cite{DV_crossing} and below). 

Another instructive feature of (\ref{R_ising}) is the presence of a pole at $\theta=iu$, with $u$ satisfying $\sin u=1-\frac{h^2}{2m}$. For $h^2<2m$ this pole is in the physical strip and corresponds to a boundary bound state, i.e. to a bound state formed by the boundary and the bulk particle which has to be interpreted as a stable excited state of the boundary; the difference between the energy of the excited boundary and that of the ordinary boundary is given by the energy of the particle evaluated at $\theta=iu$, i.e. by
\EQ
e_{B'}-e_B=m\cos u\,.
\label{boundary_binding}
\EN

\subsection{Interface in a wedge}
\label{wedge}
Interfacial phenomena at boundaries are extensively studied for their relevance in applications involving adsorption of fluids on substrates (see e.g. \cite{deGennes_wetting,DPR,BEIMR}). In this and in the next subsection we show how field theory provides general and exact results for universal properties in the two-dimensional case \cite{DS_wetting,DS_wedge}. We start from the half-plane with the boundary located at $x=0$ and consider the scaling limit below $T_c$ in which two or more phases coexist; as usual the kinks $K_{ab}$ provide the elementary excitations in the particle description. We denote by $B_a$ a boundary condition corresponding to a boundary field favoring phase $a$ in the bulk. We can switch from boundary condition $B_a$ for $y>y_0$ to boundary condition $B_b$ for $y<y_0$ inserting at $(x,y)=(0,y_0)$ a boundary condition changing field $\mu_{ab}$ with the following defining properties. First of all it has to couple to kink states connecting the vacua $|0_a\rangle_0$ and $|0_b\rangle_0$, where the subscript $0$ indicates that we consider the quantum system defined for $x\geq 0$. The simplest matrix element of $\mu_{ab}$ is
\EQ
{}_0\langle 0_a|\mu_{ab}(0,y)|K_{ba}(\theta)\rangle_0\equiv e^{-my\cosh\theta}{\cal F}^\mu_0(\theta)\,,
\label{bff}
\EN
where ${\cal F}^\mu_0(\theta)$ gives the amplitude for the emission of a kink from the boundary condition changing point. For the time being we want to exclude the possibility that the kink remains stuck on the boundary, and then we require
${\cal F}_0^\mu(0)=0$ or, on general analyticity grounds, ${\cal F}_0^\mu(\theta)=c\,\theta+O(\theta^2)$, with $c$ a constant.

\begin{figure}[t]
\centering
\includegraphics[width=3cm]{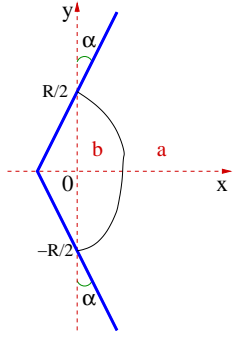}
\caption{Wedge geometry with boundary condition changing points at $(0,\pm R/2)$ and the interface running between them.}
\label{fpu_wedge}
\end{figure}

For reasons to become immediately clear we observe, recalling (\ref{rapidity}), that a rotation which brings the boundary to form an angle $\alpha$ with the vertical axis corresponds in real time to a relativistic transformation which shifts rapidities by $-i\alpha$, so that the kink emission amplitude ${\cal F}^\mu_\alpha$ in the rotated frame is related to that in the original frame as
\EQ
{\cal F}^\mu_0(\theta)={\cal F}^\mu_\alpha(\theta-i\alpha)\,;
\label{Fcovariance}
\EN
in particular
\EQ
{\cal F}_\alpha^\mu(\theta)\sim c(\theta+i\alpha)\,,\hspace{1cm}|\theta|,\,|\alpha|\ll 1\,.
\label{fmu}
\EN
With this information we can consider, instead of the half-plane, the more general wedge geometry of Fig.~\ref{fpu_wedge}. The points $(0,\pm R/2)$ are boundary condition changing points, such that a boundary field favors phase $a$ (resp. $b$) for $|y|>R/2$ (resp. $|y|<R/2$) on the wedge. With these boundary conditions the order parameter in the wedge, that we denote by $\langle\sigma(x,y)\rangle_{W_{aba}}$, tends for $x\to\infty$ to $\langle\sigma\rangle_a$ if $R$ is finite, and to $\langle\sigma\rangle_b$ if $R$ is infinite. For $mR$ large we then expect an interface running between the points $(0,\pm R/2)$, separating an inner phase $b$ from an outer phase $a$, and whose average mid-point distance from the wedge diverges with $R$. 

The theory confirms these expectations as follows. The order parameter for $|y|<R/2$ reads
\EQ
\langle \sigma(x,y) \rangle_{W_{aba}} = \frac{{}_\alpha\langle 0_{a} \vert \mu_{ab}(0,\frac{R}{2})\sigma(x,y)\mu_{ba}(0,-\frac{R}{2}) \vert 0_{a} \rangle_{-\alpha}}{{Z}_{W_{aba}}}\,, 
\label{op}
\EN
where the subscripts $\pm\alpha$ indicate the different rotations performed for positive and negative $y$, and 
\bea
{Z}_{W_{aba}} &=& {}_\alpha\langle 0_{a} \vert \mu_{ab}(0,R/2)\mu_{ba}(0,-R/2) \vert 0_{a} \rangle_{-\alpha}\,\nonumber\\
&\sim & \int_0^\infty\frac{\rd \theta}{2\pi}\,\mathcal{F}^\mu_\alpha(\theta)\mathcal{F}^\mu_{-\alpha}(\theta)\,\text{e}^{-mR(1+\frac{\theta^2}{2})}
\sim \frac{c^{2}\,\text{e}^{-mR}}{2\sqrt{2\pi}(mR)^{3/2}} \, (1 + mR \,\alpha^2)\,;
\label{partition02}
\eea
here we considered the limit $mR$ large which projects on the lightest (single-kink) intermediate state, and $\alpha$ small in order to use (\ref{fmu}). Similarly we have
\bea \nonumber
\langle \sigma(x,y) \rangle_{W_{aba}} &\sim &\frac{e^{-mR}}{{Z}_{W_{aba}}} \int_{-\infty}^{+\infty}\frac{\rd \theta_{1} \rd \theta_{2}}{(2\pi)^{2}}
\text{e}^{-\frac{m}{2}[(\frac{R}{2}-y)\theta_{1}^2+(\frac{R}{2}+y)\theta_{2}^2]+imx(\theta_{1}-\theta_{2})}\\
& \times & \mathcal{F}^\mu_{-\alpha}(\theta_{2})\, \langle K_{ab}(\theta_{1})\vert \sigma(0,0) \vert K_{ba}(\theta_{2}) \rangle\, \mathcal{F}^\mu_{\alpha}(\theta_{1})\,,
\label{op1}
\eea
where we evaluate the order parameter field on bulk states implying that the boundary condition changing fields account for the leading boundary effects at large $R$. Recalling (\ref{F2conn}) and (\ref{kappa_chi}) we obtain \cite{DS_wedge}
\EQ
\frac{\partial_{x}\langle \sigma(x,y)\rangle_{W_{aba}}}{\langle\sigma\rangle_a-\langle\sigma\rangle_b}\sim 8\sqrt{2}\left( \frac{m}{R}\right)^{\frac{3}{2}} \frac{\left(x+\frac{R\alpha}{2}\right)^{2}-(\alpha y)^{2}}{\sqrt{\pi}\,\kappa^{3}(1+mR\alpha^{2})}\,\text{e}^{-\chi^{2}},
\label{dop}
\EN
and, integrating back over $x$ with the condition $\langle\sigma(+\infty,y)\rangle_{W_{aba}}=\langle\sigma\rangle_{a}$,
\EQ
\langle \sigma(x,y) \rangle_{W_{aba}}\sim\langle\sigma\rangle_{b}+[\langle\sigma\rangle_a-\langle\sigma\rangle_b]\Biggl[\text{erf}(\chi)
-\frac{2}{\sqrt{\pi}}\,\frac{\chi + \sqrt{2mR}\,\frac{\alpha}{\kappa}}{1+mR\alpha^{2}}\,\text{e}^{-\chi^{2}} \Biggr];
\label{op2}
\EN
for $\alpha=y=0$ and $\langle\sigma\rangle_{a}=-\langle\sigma\rangle_{b}$ this result coincides with that obtained in \cite{Abraham_wetting} from the exact lattice solution of the Ising model on the half plane. Proceeding as for phase separation in the strip, we can write 
\EQ
\langle \sigma(x,y) \rangle_{W_{aba}}\sim\langle \sigma \rangle_{a} \int_{\tilde{x}}^{x} \rd v \, p(v;y) + \langle \sigma \rangle_{b} \int_{x}^{\infty} \rd v \, p(v;y)\,,
\EN
where $p(v;y)$ is the probability that the interface passes in the interval $(v,v+dv)$ on the line of constant ordinate $y$, which intersects the wedge at $x=\tilde{x}(y)$; it follows that $p(x;y)$ coincides with (\ref{dop}). The requirement $\int_{\tilde{x}}^{\infty}\rd x\,p(x;y)\approx 1$ leads to $\sqrt{mR}\,\alpha\ll 1$. Notice that (\ref{dop}) shows that $p(x;y)$ vanishes for $|y|=\frac{x}{\alpha}+\frac{R}{2}$, which for the present case of small $\alpha$ are the coordinates of the wedge ($x\geq-R\alpha/2$); hence, the properties (\ref{Fcovariance}), (\ref{fmu}) that we identified in momentum space indeed lead to an impenetrable wedge in coordinate space. A plot of $p(x;y)$ is shown in Fig.~\ref{wedge_pp}. 

\begin{figure}[t]
\centering
\includegraphics[width=7cm]{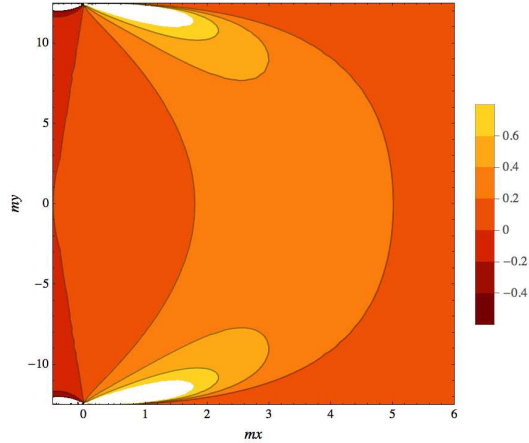}
\caption{Passage probability density $p(x;y)/m$ for the interface from (\ref{dop}) with $mR=25$, $\alpha=0.04$. The leftmost contour line corresponds to $p(x;y)=0$, and then to the wedge. As expected, the passage probability density diverges as the pinning points $(0,\pm R/2)$ are approached.}
\label{wedge_pp}
\end{figure}

\subsection{Boundary wetting and wedge filling}
\label{wetting}
In deriving the results (\ref{dop}), (\ref{op2}) we considered a regime in which the kink (i.e. the interface) does not stick to the boundary. On the other hand, considering first the case of the half-plane ($\alpha=0$), we saw in section~\ref{boundary} that in some range of the boundary parameters the particle may form with the boundary a stable bound state with energy $e_{B'}$ specified by (\ref{boundary_binding}), i.e. smaller than the energy $e_B+m\cosh\theta$ of the unbound boundary-particle system. In the present case the particle is a kink and such a boundary bound state, let us denote it $|0'_a\rangle_0$, corresponds to the phase $b$ forming a thin layer adsorbed on the boundary (Fig.~\ref{fpu_bbs}). When this state is present it dominates the spectral decomposition of (\ref{op}) at large $mR$ and produces an order parameter equal to $\langle\sigma\rangle_a$ for $mx\gg 1$, no matter how large $R$ is \cite{DS_wetting}. We know from (\ref{excess}) that $m$ is the interfacial tension between the phases $a$ and $b$; on the other hand, $e_B$ corresponds to the interfacial tension between the boundary and phase $b$, and $e'_B$ that between the excited boundary and phase $a$ (Fig.~\ref{fpu_bbs}). It follows that (\ref{boundary_binding}) amounts to the condition which in the phenomenological theory of boundary wetting (see e.g. \cite{BEIMR}) determines the equilibrium of a drop of phase $b$ between the boundary and phase $a$. Moreover, the bound state parameter $u$ is identified with the {\it contact angle} of the phenomenological theory, i.e. with the angle that the drop forms with the boundary at the contact point. The wetting transition is phenomenologically identified with the vanishing of the contact angle, when the drop spreads on the whole boundary; in field theory it corresponds to the unbinding of the kink from the boundary which takes place at $u=0$ \cite{DS_wetting}. 

For fixed values of the boundary parameters, $u$ is a function of the temperature, which is related to the bulk correlation length $\xi\propto m^{-1}\propto (T_c-T)^{-\nu}$, so that the condition $u(T_0)=0$ defines the wetting transition temperature $T_0<T_c$ for the flat boundary. For the Ising model with a boundary magnetic field $h$ discussed in section~\ref{boundary} we have $1-\sin u=\frac{h^2}{2m}=\frac{T_c-T_0(h)}{T_c-T}$; the last equality follows from $\nu_{Ising}=1$ and holds in the scaling limit.

\begin{figure}[t]
\centering
\includegraphics[width=1.3cm]{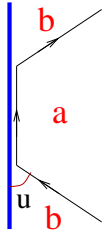}
\caption{When the kink rapidity approaches the resonance value $iu$ the kink and the boundary form a boundary bound state in which phase $b$ is adsorbed on the boundary as a thin layer; $u$ becomes the contact angle of the phenomenological wetting theory.}
\label{fpu_bbs}
\end{figure}

In order to consider the case of the wedge ($\alpha>0$), we recall first of all that matrix elements of fields on particle states inherit from the $S$-matrix the poles corresponding to bound states. For the matrix element (\ref{bff}) the pole induced by the boundary bound state takes the form
\EQ
{\cal F}_0^\mu(\theta)={}_0\langle 0_a|\mu_{ab}(0,0)|K_{ba}(\theta)\rangle_0
\sim\frac{i\Gamma_{KB}^{B'}}{\theta-iu}\,{}_0\langle 0_a|\mu_{ab}(0,0)|0'_a\rangle_0\,,\hspace{1cm}\theta\sim iu\,,
\label{bff_pole}
\EN
where $\Gamma_{KB}^{B'}$ is the kink--boundary--excited-boundary coupling.
It then follows from (\ref{Fcovariance}) that the boundary bound state pole of ${\cal F}_\alpha^\mu(\theta)$ is located at $\theta=i(u-\alpha)$. Since the kink energy is always $m\cosh\theta$ and $\theta=0$ remains the scattering threshold, the wedge wetting (or {\it filling}) transition temperature $T_\alpha$ is determined by the condition
\EQ
u(T_\alpha)=\alpha\,.
\label{filling}
\EN
For $u<\alpha$ the pole is located at Im\,$\theta<0$, namely outside the physical strip Im\,$\theta\in(0,\pi)$ allowed for bound states; in such a case the kink is unbounded and phase $b$ fills the wedge. The condition (\ref{filling}), which follows here from the exact field theoretical analysis, is known phenomenologically \cite{Hauge} and provides the basic illustration of how the adsorption properties are affected by the geometry of the substrate \cite{RDN,RP}. It relates the filling transition temperature in the wedge to the contact angle $u(T)$ for the flat boundary, a connection known as {\it wedge covariance} \cite{APW,RP2}; field theory shows that wedge covariance originates from the relativistic covariance expressed by (\ref{Fcovariance}) \cite{DS_wedge}.

\section{Off-critical crossing clusters}
\label{nearcritical}
Let us consider random percolation on a rectangle of sides $R$ and $L$ (Fig.~\ref{fpu_cross_cluster}). We call vertically crossing cluster a cluster touching both horizontal sides of the rectangle, and vertical crossing probability $P_v$ the probability that at least one such cluster is present. In the scaling limit $P_v$ is a universal function of $R/\xi$ and $L/\xi$, where $\xi$ is the bulk correlation length, proportional to the average linear cluster size. Crossing probabilities have been originally studied at the critical point (occupation probability $p=p_c$) \cite{LPPS-A}, where $\xi=\infty$ and $P_v$ becomes a function of a single variable, the aspect ratio $R/L$. For this case Cardy was able to determine $P_v$ exactly mapping it to a four-point function of boundary condition changing fields associated to the corners of the rectangle; since these fields are degenerate, in the sense explained in section~\ref{basic}, $P_v$ becomes the solution of a differential equation of hypergeometric type \cite{Cardy92}, a result which later has also been obtained through mathematically rigorous methods \cite{S.Smirnov}.

\begin{figure}[t]
\begin{center}
\includegraphics[height=3cm]{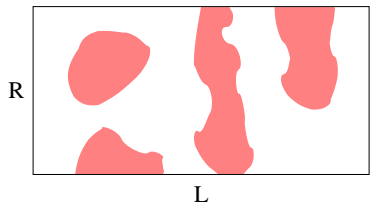}
\caption{A cluster configuration on the rectangle including a vertically crossing cluster.}
\label{fpu_cross_cluster}
\end{center}
\end{figure}

Here we discuss the near-critical case and show how, building on ideas introduced in the last sections, $P_v$ can be determined exactly in the limit in which the sides of the rectangle are much larger than $\xi$ \cite{DV_crossing}. As we regularly did to deal with percolation, we start with a mapping to the $q\to 1$ Potts model. We recall that the factor $q$ to the number of clusters in the cluster decomposition (\ref{FK}) of the Potts partition function is due to the fact that each cluster comes in $q$ colors. When putting the model on a rectangle, we have to take into account the effect of boundary conditions. Let us call $Z_{ud}^{lr}$ the Potts partition function on the rectangle with boundary conditions $u$, $d$, $l$ and $r$ on the upper, lower, left and right boundary, respectively; we will then denote by $f$ the free boundary condition, and by $a$ the boundary condition in which the spins are fixed to a color $a$. Then, for $a\neq b$, $Z_{ab}^{ff}$ does not receive a contribution from configurations containing vertically crossing clusters. On the other hand, as $q\to 1$ the weight of a configuration becomes its probability, and one has
\EQ
P_v=1-\lim_{q\to 1}Z_{ab}^{ff}\,,
\label{p_z}
\EN
a relation already used in \cite{Cardy92} for the study of the critical case.
In order to express $Z_{ud}^{lr}$ away from criticality we consider first the limit $L\to\infty$ and use the boundary state technique of section~\ref{order}. We then write
\EQ
Z_{ud}^{lr}=\langle B_u|\text{e}^{-RH}|B_d\rangle_{l,r}\,,\hspace{1cm}L\to\infty\,,
\label{Z_ud}
\EN
where now $|B_{d,u}\rangle$ are boundary states corresponding to uniform boundary conditions (fixed or free), and the vertical boundary conditions at infinity have the role of selecting the states which can propagate between the horizontal boundaries. Consider the case $T<T_c$ in which there are $q$ degenerate vacua $|0_a\rangle$ and the bulk elementary excitations are the kinks $K_{ab}$ with mass $m$. Then fixed and free boundary states expand as\footnote{See \cite{DS_casimir} for the expansion of boundary states in the basis of high-temperature excitations.}
\bea
|B_a\rangle &=& |0_a\rangle+\int\frac{d\theta}{2\pi}f(\theta)|K_{ab}(-\theta)K_{ba}(\theta)\rangle+\ldots\,,
\label{B_a}\\
|B_f\rangle &=& \sum_a\left[|0_a\rangle+\tilde{g}\sum_{b\neq a}|K_{ab}(0)\rangle+\ldots\right]\,,
\label{B_f}
\eea
respectively, where, due to translation invariance, all the states appearing in the expansion have zero total momentum, as originally illustrated in \cite{GZ}. Then we have, in particular, 
\EQ
[Z_{ff}^{ab}]^+=\langle B_f|\text{e}^{-RH}|B_f\rangle_{a,b}^+=\tilde{g}^2mL\,e^{-mR}+O(e^{-2mR})\,,
\label{Z_ff}
\EN
where the vertical boundary conditions $a\neq b$ forbid vacuum contributions, and we used $\langle K_{ab}(0)|K_{bc}(0)\rangle=2\pi\delta_{ac}\delta(0)=\delta_{ac}mL$; we see how the $L$-dependence is reintroduced through the volume regularization of the delta function. The superscript $+$ in (\ref{Z_ff}) recalls that the quantity is computed in the scaling limit below $T_c$ in the Potts model, which corresponds to the scaling limit above $p_c$ in percolation; we will use the superscript $-$ for the scaling limit below $p_c$. It is not difficult to see that there is a one-to-one correspondence between lattice Potts configurations with vertical crossings and configurations without horizontal crossings on the dual lattice \cite{Wu}; then Potts high-low temperature duality leads to \cite{DV_crossing}
\bea
P_v^- &=& 1-P_h^+=[Z_{ff}^{ab}]^+_{q=1}\sim A\,mL\,e^{-mR}\,,
\label{hv1}\\
P_v^+ &=& 1-P_h^-=1-P_v^-|_{R\leftrightarrow L}\sim 1-A\,mR\,e^{-mL}\,.
\label{hv2}
\eea
where $P_h$ is the horizontal crossing probability, $A\equiv\tilde{g}^2|_{q=1}$, and the results hold for $mR$ and $mL$ both much larger than 1. Recalling (\ref{connectivity}) we have $m=1/\xi$ below $p_c$ and $m=1/2\xi$ above $p_c$, with $\xi\propto|p-p_c|^{-4/3}$ the correlation length defined by the exponential decay of the bulk two-point connectivity. 

Recall now Fig.~\ref{fpu_boundary}b, where a pair of particles with rapidities $\theta$ and $-\theta$ is emitted from the boundary with emission amplitude $K(\theta)$. If the rapidity difference $2\theta$ coincides with a value $iu$ for which the two particles can form a bound state, $K$ will have a pole of the form $\frac{ig\Gamma}{2\theta-iu}$ \cite{GZ}, where $\Gamma$ is the three-particle coupling (Fig.~\ref{fpu_pole}) and $g$ the boundary-particle coupling. Clearly the latter is related to $\tilde{g}$ in (\ref{B_f}) \cite{GZ}, the precise relation being $\tilde{g}=g/2$ \cite{DPTW00,BPT07}. On the other hand, the Potts model on the half-plane with fixed or free boundary conditions is integrable in the scattering framework of section~\ref{boundary} \cite{Chim}, and this leads to \cite{DV_crossing}
\EQ
A=\frac{3-\sqrt{3}}{2}\,.
\EN

Notice, recalling (\ref{FK}), that $\lim_{q\to 1}\partial_q\ln Z_{ff}^{ff}$ gives the mean number of clusters. The same operation on $Z_{aa}^{ff}$ gives the mean number of clusters which do not touch the horizontal sides of the rectangle, since the others are forced to have color $a$ and to not contribute a factor $q$ to the partition function \cite{Cardy00}. Similarly, one can count also the clusters touching only the upper or lower side, and arrives to the expression 
\EQ
\bar{N}_v=\lim_{q\rightarrow 1}\partial_q\ln\frac{Z_{ff}^{ff}Z_{aa}^{ff}}{Z_{a f}^{ff}Z_{fa}^{ff}}\,
\label{Nc}
\EN
for the mean number of vertically crossing clusters. When using (\ref{B_a}) and (\ref{B_f}) to evaluate the partition functions above $p_c$ in the limit of large $L$ and $R$, the finite size corrections\footnote{Notice that (\ref{Z_ud}) determines the partition function up to a bulk contribution which does not affect our results for crossing clusters.} are dominated by the single-kink contribution coming from $Z_{ff}^{ff}$, and we obtain \cite{DV_crossing}
\EQ
\bar{N}_v^+\sim \lim_{q\rightarrow 1}\partial_q\ln {Z_{ff}^{ff}}\sim \lim_{q\rightarrow 1}\partial_q\ln[q+q(q-1)mL\tilde{g}^2e^{-mR}]\sim 1+A\,mL\,e^{-mR};
\EN
the first term is produced by the vacuum contributions and shows that above $p_c$ there is a single cluster which becomes infinite in the limit $R\to\infty$. Below $p_c$, on the other hand, vertically crossing clusters become rare as $R$ becomes large, and we have $P_v^-\equiv\text{Prob}(N_v^->0)\sim\text{Prob}(N_v^-=1)\sim\bar{N}_v^-$. 

Next-to-leading corrections are computed in \cite{DV_crossing}, where the comparison with numerical data can also be found. In principle, integrability allows an exact evaluation of partition functions on strips also for $R$ not large, through the boundary version \cite{LMSS} of the thermodynamic Bethe ansatz; at present, however, the thermodynamic Bethe ansatz for the Potts model with continuous values of $q$ is unavailable even in absence of boundaries (see  \cite{DPT}).

\section{Conclusion} 
In this article we illustrated how field theory leads to the exact description of universal properties of two-dimensional statistical systems. We tried to convey in simple terms the origin of the specific effectiveness of field theory in two dimensions, and showed its wide applicability, which allows one to treat within the same theoretical framework problems such as percolation and wetting, usually considered as uncorrelated. For the critical case, we stressed the role of conformal field theory, but also pointed out that of the particle description in the study of bulk properties of clusters and walks. The particle formalism becomes preponderant away from criticality, where conformal symmetry is lost but is replaced, in most of the interesting cases, by integrability, i.e. exact solvability of the particle dynamics. We saw that integrability is crucial in the non-perturbative study of near-critical singular behavior and of crossover phenomena. It plays a somewhat different role in problems, such as phase separation, in which the system is observed on scales much larger than the correlation length. In this case the order parameter profile and the fluctuations of the interface are determined exactly by general low-energy properties of two-dimensional field theory, but in general the analysis requires some information on bound states, which is provided by integrability and determines the wetting properties, both in the bulk and at boundaries. 

We also exhibited the important role played by the fact that in two dimensions cluster boundaries and interfaces can be associated to particle trajectories in imaginary time. Remarkably this leads, in particular, to a manifestation in real space of properties of the underlying particle description. So, the contact angle of phenomenological wetting theory corresponds to the position of a bound state pole, the dependence on the opening angle for wetting in a wedge (wedge covariance) reflects relativistic covariance, the asymptotic amplitude of crossing probabilities in near-critical percolation is determined by particle-boundary scattering.
The fact that these and other results have been obtained in the last few years suggests that additional interesting developments can reasonably be expected for the future.


\vspace{1cm} \noindent \textbf{Acknowledgments.} I thank for hospitality the Galileo Galilei Institute for Theoretical Physics in Arcetri, Florence, where this article was completed and related lectures were given during the winter school "Statistical Field Theories 2015".


\begin{thebibliography}{99}
\bibitem{LL} L.D. Landau and E.M. Lifshitz, Statistical Physics, part 1, 3rd edn, Elsevier, 1980. 
\bibitem{Potts} R.B. Potts, Proc. Cambr. Phil. Soc. 48 (1952) 106.
\bibitem{Wu} F.Y. Wu, Rev. Mod. Phys. 54 (1982) 235.
\bibitem{FK}  C.M. Fortuin and P.W. Kasteleyn, J. Phys. Soc. Jpn. Suppl. 26 (1969) 11; Physica 57 (1972) 536.
\bibitem{SA} D. Stauffer and A. Aharony, Introduction to Percolation Theory, 2nd edn, Taylor \& Francis, London, 1992.
\bibitem{Grimmett} G. Grimmett, Percolation, 2nd edn, Springer, Berlin, 1999.
\bibitem{Cardy_book} J. Cardy, Scaling and renormalization in statistical physics, Cambridge, 1996.
\bibitem{deGennes_walks} P.G. de Gennes, Phys. Lett. A 38 (1972) 339.
\bibitem{WK} K.G. Wilson and J. Kogut, Physics Reports 12 (1974) 75.
\bibitem{Weinberg} S. Weinberg, The theory of quanum fields, vol. 2, Cambridge, 1996.
\bibitem{ELOP} R.J. Eden, P.V. Landshoff, D.I. Olive and J.C. Polkinghorne, The analytic S-matrix, Cambridge, 1966.
\bibitem{BPZ}  A.A. Belavin, A.M. Polyakov and A.B. Zamolodchikov, Nucl. Phys. B 241 (1984) 333.
\bibitem{DfMS} P. Di Francesco, P. Mathieu and D. Senechal, Conformal field theory, Springer-Verlag, New York, 1997.
\bibitem{Z_cth} A.B. Zamolodchikov, JETP Lett. 43 (1986) 730.
\bibitem{DF} Vl.S. Dotsenko and V.A. Fateev, Nucl. Phys. B 240 (1984) 312.
\bibitem{Coleman} S. Coleman, Phys. Rev. D 11 (1975) 2088.
\bibitem{Mandelstam} S. Mandelstam, Phys. Rev. D 11 (1975) 3026.
\bibitem{ZZ_cft} A.B. Zamolodchikov and Al.B. Zamolodchikov, Sov. Sci. Rev. A: Phys. 10 (1989) 269-433.
\bibitem{KT} J.M. Kosterlitz and D.J. Thouless, J. Phys. C 6 (1973) 1181.
\bibitem{MW} N.D. Mermin and H. Wagner, Phys. Rev. Lett. 17 (1966) 1133. 
\bibitem{Hohenberg} P.C. Hohenberg, Phys. Rev. 158 (1967) 383.
\bibitem{Coleman_goldstone} S. Coleman, Comm. Math. Phys. 31 (1973) 259. 
\bibitem{Kaufman} B. Kaufman, Phys. Rev. 76 (1949) 1244.
\bibitem{Onsager} L. Onsager, Phys. Rev. 65 (1944) 117.
\bibitem{Yang} C.N. Yang, Phys. Rev. 85 (1952) 808.
\bibitem{FK_paraf} E. Fradkin and L.P. Kadanoff, Nucl. Phys. B 170 (1980) 1.
\bibitem{FZ_paraf} V.A. Fateev and A.B. Zamolodchikov, Sov. Phys. JETP 62 (1985) 215.
\bibitem{paraf} G. Delfino, Ann. Phys. 333 (2013) 1.
\bibitem{FQS} D. Friedan, Z. Qiu and S. Shenker, Phys. Rev. Lett. 52 (1984) 1575.
\bibitem{Zamo_multicritical} A.B. Zamolodchikov, Sov. J. Nucl. Phys. 44 (1986) 529.
\bibitem{Huse} D.A. Huse, Phys. Rev. B 30 (1984) 3908.
\bibitem{Cardy_modular} J.L. Cardy, Nucl. Phys. B 270 (1986) 186.
\bibitem{CIZ} A. Cappelli, C. Itzykson and J.B. Zuber, Nucl. Phys. B 280 (1987) 445.
\bibitem{Dotsenko} V.S. Dotsenko, Nucl. Phys. B 235 (1984) 54.
\bibitem{Gurarie} V. Gurarie, Nucl. Phys. B 410 (1993) 535.
\bibitem{GJRSV} A.M. Gainutdinov, J.L. Jacobsen, N. Read, H. Saleur and R. Vasseur, J. Phys. A 46 (2013) 494012.
\bibitem{DMM} G. Delfino, P. Mosconi and G. Mussardo, J. Phys. A 36 (2003) L1.
\bibitem{Nienhuis} B. Nienhuis, in C. Domb and J.L. Lebowitz (eds.), Phase transitions and critical phenomena, vol. 11, p. 1, Academic Press, London, 1987.
\bibitem{selfavoiding} A.B. Zamolodchikov, Mod. Phys. Lett. A 6 (1991) 1807.
\bibitem{CZ} L. Chim and A.B. Zamolodchikov, Int. J. Mod. Phys. A 7 (1992) 5317.
\bibitem{Baxter} R.J. Baxter, Exactly Solved Models of Statistical Mechanics, Academic Press, London, 1982.
\bibitem{Taniguchi} A.B. Zamolodchikov, Advanced Studies in Pure Mathematics 19 (1989) 641; Int. J. Mod. Phys. A 4 (1989) 4235.
\bibitem{VJS} E. Vernier, J.L. Jacobsen and H. Saleur, J. Phys. A 47 (2014) 285202.
\bibitem{SW} R. Shankar and E. Witten, Phys. Rev. D 17 (1978) 2134.
\bibitem{ZZ} A.B. Zamolodchikov and Al.B. Zamolodchikov, Ann. Phys. 120 (1979) 
253.
\bibitem{ising_duality} H.A. Kramers and G.H. Wannier, Phys. Rev. 60 (1941) 252.
\bibitem{KS} R. Koberle and J.A. Swieca Phys. Lett. B 86 (1979) 209.
\bibitem{Zamo_3potts} A.B. Zamolodchikov, Int. J. Mod. Phys. A3 (1988) 743.
\bibitem{TBA} Al.B. Zamolodchikov, Nucl. Phys. B 342 (1990) 695. 
\bibitem{review} G. Delfino, J. Phys. A 37 (2004) R45.
\bibitem{KB} L.P. Kadanoff and A.C. Brown, Ann. Phys. 121 (1979) 318.
\bibitem{Kadanoff} L.P. Kadanoff, Phys. Rev. B 22 (1980) 1405.
\bibitem{AT} G. Delfino, Phys. Lett. B 450 (1999) 196.
\bibitem{ATratios} G. Delfino and P. Grinza, Nucl. Phys. B 682 (2004) 521.
\bibitem{LW} E.H. Lieb and F.Y. Wu, in: C. Domb, M.S. Green (Eds.), Phase transitions and critical phenomena, Vol. 1, p.~332, Academic Press, London, 1972.
\bibitem{Baxter_AF} R.J. Baxter, Proc. Roy. Soc. London 383 (1982) 43. 
\bibitem{AF} G. Delfino, in: Statistical Field Theories, p.~3, A. Cappelli and G. Mussardo (Eds.), Kluwer Academic, Amsterdam, 2002; J. Phys. A 34 (2001) L311.
\bibitem{CJS} J. Cardy, J. Jacobsen and A.D. Sokal, J. Stat. Phys. 105 (2001) 25.
\bibitem{HR} D.A. Huse and A.D. Rutenberg, Phys. Rev. B 45 (1992) 7536. 
\bibitem{KW} M. Karowski, P. Weisz, Nucl. Phys. B 139 (1978) 455.
\bibitem{Smirnov} F.A. Smirnov, Form factors in completely integrable models of quantum field theory, World Scientific, 1992.
\bibitem{DV_ps} G. Delfino and J. Viti, J. Stat. Mech. (2012) P10009.
\bibitem{immf} G. Delfino and G. Mussardo, Nucl. Phys. B 455 (1995) 724.
\bibitem{D09} G. Delfino, Nucl. Phys. B 807 (2009) 455.
\bibitem{BKW} B. Berg, M. Karowski, P. Weisz, Phys. Rev. D 19 (1979) 2477. 
\bibitem{DSC} G. Delfino, P. Simonetti and J.L. Cardy, Phys. Lett. B 387 
(1996) 327. 
\bibitem{WMcTB}T.T. Wu, B.M. McCoy, C.A. Tracy and E. Barouch, Phys. Rev. B 13 
(1976) 316.
\bibitem{ising_ratios} G. Delfino, Phys. Lett. B 419 (1998) 291; Erratum, Phys. Lett. B 518 (2001) 330.
\bibitem{DC} G. Delfino and J. Cardy, Nucl. Phys. B 519 (1998) 551.
\bibitem{DCq4} G. Delfino and J. Cardy, Phys. Lett. B 483 (2000) 303.
\bibitem{DVC} G. Delfino, J. Viti and J. Cardy, J. Phys. A 43 (2010) 152001.
\bibitem{JZ} I. Jensen and R. Ziff, Phys. Rev. E 74 (2006) 020101(R).
\bibitem{EG} I.G. Enting and A.J. Guttmann, Physica A 321 (2003) 90.
\bibitem{SBB} L.N. Shchur, B. Berche and P. Butera, Phys. Rev. B 77 (2008) 144410.
\bibitem{SJ} L.N. Shchur and W. Janke, Nucl. Phys. B 840 (2010) 491. 
\bibitem{Lukyanov} S. Lukyanov, Mod. Phys. Lett. A12 (1997) 2543.
\bibitem{PHA} V. Privman, P.C. Hohenberg and A. Aharony, in: Phase transition and critical phenomena, Vol. 14, C. Domb and J.L. Lebowitz eds., Academic Press, 1991.
\bibitem{Ziff} R.M. Ziff, in: {Proceedings of the 24th workshop on computer simulation studies in condensed matter physics}, D.P. Landau ed.,p. 106, 2011.
\bibitem{Alyosha} Al.B. Zamolodchikov, Nucl. Phys. B 358 (1991) 524.
\bibitem{DMS_flow} G. Delfino, G. Mussardo and P. Simonetti, Phys. Rev. D 51 (1995) 6620.
\bibitem{CNPR} A. Coniglio, C. Nappi, F. Peruggi and L. Russo, J. Phys. A 10 (1977) 205.
\bibitem{Murata} K.K. Murata, J. Phys. A 12 (1979) 81.
\bibitem{CK} A. Coniglio and W. Klein, J. Phys. A 13 (1980) 2775.
\bibitem{NBRS} B. Nienhuis, A.N. Berker, E.K. Riedel and M. Shick, Phys. Rev. Lett. 43 (1979) 737.
\bibitem{isingperc} G. Delfino, Nucl. Phys. B 818 (2009) 196.
\bibitem{FSZ} P. Fendley, H. Saleur and Al.B. Zamolodchikov, Int. J. Mod. Phys. A 8 (1993) 5751.
\bibitem{DS_double} G. Delfino and A. Squarcini, Ann. Phys. 342 (2014) 171.
\bibitem{Abraham_81} D.B. Abraham, Phys. Rev. Lett. 47 (1981) 545.
\bibitem{Abraham} D.B. Abraham, in: C. Domb, J.L. Lebowitz (Eds.), Phase transitions and critical phenomena, Vol. 10, p.~1, Academic Press, London, 1986.
\bibitem{D_dilute} G. Delfino, Nucl. Phys. B 554 (1999) 537.
\bibitem{AU} D.B.~Abraham and P.J.~Upton, {Phys. Rev. Lett.} 70 (1993) 1567.
\bibitem{KoF} L.-F. Ko and M.E.~Fisher, {J. Stat. Phys.} 58 (1990) 249.
\bibitem{Fisher_vicious} M.E.~Fisher, J. Stat. Phys. 34 (1984) 667.
\bibitem{Schramm} O. Schramm, Israel J. Math. 118 (2000) 221; Electronic Comm. Probab. 8 (2001) paper no. 12.
\bibitem{BB} M. Bauer and D. Bernard, Phys. Lett. B 543 (2002) 135; Comm. Math. Phys. 239 (2003) 493.
\bibitem{Cardy_sle} J. Cardy, Ann. Phys. 318 (2005) 81.
\bibitem{BBreport} M. Bauer and D. Bernard, Phys. Rep. 432 (2006) 115.
\bibitem{GC} A. Gamsa and J. Cardy, J. Stat. Mech. (2007) P08020.
\bibitem{DJS} J. Dubail, J.L. Jacobsen and H. Saleur, J. Phys. A 43 (2010) 482002; J. Stat. Mech. P12026 (2010).
\bibitem{Dietrich} S. Dietrich, in: C. Domb and J.L. Lebowitz (Eds.), Phase Transitions and Critical Phenomena, Vol. 12, p.~1, Academic Press, London, 1988.
\bibitem{conf1} G. Delfino, G. Mussardo and P. Simonetti, Nucl. Phys. B 473 (1996) 469.
\bibitem{conf2} G. Delfino and G. Mussardo, Nucl. Phys. B 516 (1998) 675.
\bibitem{McW} B.M. McCoy and T.T. Wu, Phys. Rev. D 18 (1978) 1259.
\bibitem{FZspectroscopy} P. Fonseca and A.B. Zamolodchikov, J. Stat. Phys. 110 (2003) 527; arXiv:hep-th/0309228; arXiv:hep-th/0612304.
\bibitem{DGM} G. Delfino, P. Grinza and G. Mussardo, Nucl. Phys. B 737 (2006) 291.
\bibitem{DGconf} G. Delfino and P. Grinza, Nucl. Phys. B 791 (2008) 265. 
\bibitem{LTD} L. Lepori, G.Z. Toth and G. Delfino, J. Stat. Mech. (2009) P11007.
\bibitem{Rutkevich} S.B. Rutkevich, J. Phys. A 43 (2010) 235004; J. Stat. Mech. (2015) P01010.
\bibitem{Kardar} M. Kardar, Nucl. Phys. B 290 (1987) 582.
\bibitem{DD} V.S. Dotsenko and V1.S. Dotsenko, Sov. Phys. JETP Lett. 33 (1981) 37; Adv. Phys. 32 (1983) 129.
\bibitem{Zamo_spectroscopyII} A.B. Zamolodchikov, arXiv:1310.4821.
\bibitem{Binder} K. Binder, in: C. Domb and J.L. Lebowitz (Eds.), Phase Transitions and Critical Phenomena, Vol. 8, p.~1, Academic Press, London, 1983.
\bibitem{Diehl} H.W. Diehl, in: C. Domb and J.L. Lebowitz (Eds.), Phase Transitions and Critical Phenomena, Vol. 10, p.~75, Academic Press, London, 1986.
\bibitem{Cardy_bcft} J.L. Cardy, in: Encyclopedia of Mathematical Physics, Elsevier, 2006.
\bibitem{GZ} S. Ghoshal and A.B. Zamolodchikov, Int. J. Mod. Phys. A9 (1994) 3841; Erratum, ibidem A9 (1994) 4353.
\bibitem{FringK} A. Fring and R. Koberle, Nucl. Phys. B 421 (1994) 159.
\bibitem{Cardy_89} J.L. Cardy, Nucl. Phys. B 324 (1989) 581.
\bibitem{Chim} L. Chim, J. Phys. A: Math Gen. 28 (1995) 7039.
\bibitem{DV_crossing} G. Delfino and J. Viti, J. Phys. A 45 (2012) 032005.
\bibitem{deGennes_wetting} P.G. De Gennes, Rev. Mod. Phys. 57 (1985) 827.
\bibitem{DPR} S. Dietrich, M.N. Popescu and M. Rauscher, J. Phys. Condens. Matter 17 (2005) S577.
\bibitem{BEIMR} D. Bonn, J. Eggers, J. Indekeu, J. Meunier and E. Rolley, Rev. Mod. Phys. 81 (2009) 739.
\bibitem{DS_wetting} G.~Delfino and A. Squarcini, J. Stat. Mech. (2013) P05010.
\bibitem{DS_wedge} G. Delfino and A. Squarcini, Phys. Rev. Lett. 113 (2014) 066101.
\bibitem{Abraham_wetting} D.B. Abraham, Phys. Rev. Lett. 44 (1980) 1165.
\bibitem{Hauge} E.H. Hauge, Phys. Rev. A 46 (1992) 4994. 
\bibitem{RDN} K. Rejmer, S. Dietrich and M. Napiorkowski, Phys. Rev. E 60 (1999) 4027.
\bibitem{RP} C. Rasc\'on and A.O. Parry, Nature (London) 407 (2000) 986.
\bibitem{APW} D.B. Abraham, A.O. Parry and A.J. Wood, Europhys. Lett. 60 (2002) 106.
\bibitem{RP2} C. Rasc\'on and A.O. Parry, Phys. Rev. Lett. 94 (2005) 096103.
\bibitem{LPPS-A} R. Langlands, C. Pichet, P. Pouliot and Y. Saint-Aubin, J. Stat. Phys. 67 (1992) 553.
\bibitem{Cardy92} J. Cardy, J. Phys. A 25 (1992) L201.
\bibitem{S.Smirnov} S. Smirnov, C.R. Acad. Sci. Paris, t. 333, S\'erie I (2001) 339.
\bibitem{DS_casimir} G. Delfino and A. Squarcini, EPL 109 (2015) 16001.
\bibitem{DPTW00} P. Dorey, M. Pillin, R. Tateo and G. Watts, Nucl. Phys. B 594 (2000) 625.
\bibitem{BPT07} Z. Bajnok, L. Palla and G. Takacs, Nucl. Phys. B 772 (2007) 290.
\bibitem{Cardy00} J. Cardy, Phys. Rev. Lett. 84 (2000) 3507.
\bibitem{LMSS} A. Le Clair, G. Mussardo, H. Saleur and S. Skorik, Nucl. Phys. B 453 (1995) 581.
\bibitem{DPT} P. Dorey, A. Pocklington and R. Tateo, Nucl. Phys. B 661 (2003) 464.



\end{thebibliography}
\end{document}